\documentclass[12pt,a4paper,fleqn]{article}
\usepackage{tikz}
\usepackage{caption}
\usepackage{lscape}

\usepackage{fullpage}
\usepackage{amsmath}
\usepackage{amsthm}
\usepackage{amssymb}
\usepackage{amscd}
\usepackage{plain}
\usepackage{graphicx}
\usepackage{t1enc,epsfig}
\usepackage[mathscr]{eucal}
\usepackage{epstopdf}
\usepackage[german,english]{babel}
\usepackage{stmaryrd}

\usepackage{xcolor}
\usepackage{float}

\usepackage[colorlinks, allcolors=blue]{hyperref}
\usepackage[normalem]{ulem}

\numberwithin{equation}{section}

\begin{document}

\def\diy{\displaystyle}

\def\pl {\partial}

\def\rd{{\rm d}} \def\re{{\rm e}} \def\ri{{\rm i}}
\def\rt{{\rm t}} \def\rv{{\rm v}} \def\rw{{\rm w}}
\def\rx{{\rm x}} \def\ry{{\rm y}} \def\rz{{\rm z}}
\def\rD{{\rm D}}
\def\rO{{\rm O}} \def\rP{{\rm P}} \def\rV{{\rm V}}

\def\bD{{\mathbf D}}

\def\ovphi{\ov\vphi}
\def\ophi{\ov\phi}

\def\bbA{{\mathbb A}} \def\bbB{{\mathbb B}} \def\bbC{{\mathbb C}}
\def\bbD{{\mathbb D}} \def\bbE{{\mathbb E}} \def\bbF{{\mathbb F}}
\def\bbG{{\mathbb G}} \def\bbH{{\mathbb H}}
\def\bbL{{\mathbb L}} \def\bbM{{\mathbb M}} \def\bbN{{\mathbb N}}
\def\bbP{{\mathbb P}} \def\bbQ{{\mathbb Q}}
\def\bbR{{\mathbb R}} \def\bbS{{\mathbb S}} \def\bbT{{\mathbb T}}
\def\bbU{{\mathbb U}}
\def\bbV{{\mathbb V}} \def\bbW{{\mathbb W}} \def\bbZ{{\mathbb Z}}

\def\bbZd{\bbZ^d}

\def\cA{{\mathcal A}} \def\cB{{\mathcal B}} \def\cC{{\mathcal C}}
\def\cD{{\mathcal D}} \def\cE{{\mathcal E}} \def\cF{{\mathcal F}}
\def\cG{{\mathcal G}} \def\cH{{\mathcal H}} \def\cJ{{\mathcal J}}
\def\cP{{\mathcal P}} \def\cS{{\mathcal S}}  \def\cT{{\mathcal T}} 
\def\cW{{\mathcal W}}
\def\cX{{\mathcal X}} \def\cY{{\mathcal Y}} \def\cZ{{\mathcal Z}}

\def\tA{{\tt A}} \def\tB{{\tt B}}
\def\tC{{\tt C}} \def\tF{{\tt F}} \def\tH{{\tt H}}
\def\tO{{\tt O}}
\def\tP{{\tt P}} \def\tQ{{\tt Q}} \def\tR{{\tt R}}
\def\tS{{\tt S}} \def\tT{{\tt T}}

\def\tx{{\tt x}} \def\ty{{\tt y}}

\def\t0{{\tt 0}} \def\t1{{\tt 1}}

\def\be{{\mathbf e}} \def\bh{{\mathbf h}}
\def\bn{{\mathbf n}} \def\bu{{\mathbf u}}
\def\bx{{\mathbf x}} \def\by{{\mathbf y}} \def\bz{{\mathbf z}}
\def\B1{{\mathbf 1}} \def\co{\complement}

\def\fA{{\mathfrak A}} \def\fB{{\mathfrak B}} \def\fC{{\mathfrak C}}
\def\fD{{\mathfrak D}} \def\fE{{\mathfrak E}} \def\fF{{\mathfrak F}}
\def\fW{{\mathfrak W}} \def\fX{{\mathfrak X}} \def\fY{{\mathfrak Y}}
\def\fZ{{\mathfrak Z}}

\def\rA{{\rm A}}  \def\rB{{\rm B}}  \def\rC{{\rm C}}
\def\rF{{\rm  F}} \def\rM{{\rm  M}}
\def\rS{{\rm S}}
\def\rT{{\rm T}}  \def\rW{{\rm W}}

\def\ov{\overline}  \def\un{\underline}  
\def\unf{\underline f}
\def\wh{\widehat}  \def\wt{\widetilde}

\def\es {{\varnothing}}

\def\bt {\circ}
\def\cc {\circ}

\def\wt{\widetilde}

\def\wtm{{\wt m}} \def\wtn{{\wt n}} \def\wtk{{\wt k}}

\def\be{\begin{equation}}
\def\ee{\end{equation}}

\def\beal{\begin{array}{l}}
\def\beac{\begin{array}{c}}
\def\bear{\begin{array}{r}}
\def\beacl{\begin{array}{cl}}
\def\beall{\begin{array}{ll}}
\def\bealll{\begin{array}{lll}}
\def\beallll{\begin{array}{llll}}
\def\bealllll{\begin{array}{lllll}}
\def\beacr{\begin{array}{cr}}
\def\ena{\end{array}}

\def\bma{\begin{matrix}}
\def\ema{\end{matrix}}

\def\bpma{\begin{pmatrix}}
\def\epma{\end{pmatrix}}

\def\bcs{\begin{cases}}
\def\ecs{\end{cases}}

\def\diy{\displaystyle}

\def\sA{\mathscr A} \def\sB{\mathscr B} \def\sC{\mathscr C}
\def\sD{\mathscr D} \def\sE{\mathscr E}
\def\sF{\mathscr F} \def\sG{\mathscr G} 
\def\sH{\mathscr H} \def\sI{\mathscr I}
\def\sL{\mathscr L} \def\sM{\mathscr M}
\def\sN{\mathscr N} \def\sO{\mathscr O}
\def\sP{\mathscr P} \def\sR{\mathscr R}
\def\sS{\mathscr S} \def\sT{\mathscr T}
\def\sU{\mathscr U} \def\sV{\mathscr V} \def\sW{\mathscr W}
\def\sX{\mathscr X} \def\sY{\mathscr Y} \def\sZ{\mathscr Z}

\def\BZ{{\mathbf Z}}

\def\D{D}

\def\bs {\overline \phi}
\def\bbLv {{\cal E}}

\def\iy{\infty}
\def\ct{\cdot}
\def\cl{\centerline}

\def\bbL{{\mathbb L}} \def\Pf{{\mathbf Z}} \def\f{{\vphi}} \def\g{{\Gam}} \def\s{{\phi}}
\def\boeta{{\mbox{\boldmath$\eta$}}}
\def\sq{\square} \def\tr{\triangle} \def\lz{\lozenge}
\def\cre{\color{red}} \def\cbl{\color{blue}} \def\cte{\color{teal}}
\def\cl{\centerline}

\def\bmu{{\mbox{\boldmath${\mu}$}}} 
\def\bbA{{\mathbb A}} \def\bbE{{\mathbb E}} 
\def\bbH{{\mathbb H}} \def\bbN{{\mathbb N}}
\def\bbR{{\mathbb R}} \def\bbV{{\mathbb V}}  
\def\bbW{{\mathbb W}} \def\bbZ{{\mathbb Z}}
\def\cA{{\mathcal A}} \def\cC{{\mathcal C}} 
\def\cE{{\mathcal E}}
\def\cG{{\mathcal G}} \def\cP{{\mathcal P}}

\def\alp{{\alpha}} \def\bet{{\beta}} \def\gam{{\gamma}}
\def\del{{\delta}} \def\eps{{\epsilon}} \def\lam{{\lambda}} 
\def\ups{{\upsilon}}
\def\Om {\Omega}
\def\Gam{\Gamma} \def\Lam{\Lambda} 

\def\vphi{\varphi} \def\vrho{\varrho}
\def\veps{\varepsilon} \def\vpi{\varpi}

\def\balpha{{\mbox{\boldmath${\alpha}$}}}
\def\bmu{{\mbox{\boldmath${\mu}$}}}
\def\btau{{\mbox{\boldmath${\tau}$}}}
\def\bzeta{{\mbox{\boldmath${\zeta}$}}}
\def\bnu{{\mbox{\boldmath${\nu}$}}}
\def\bPhi{{\mbox{\boldmath${\Phi}$}}}

\def\tD{{\tt D}} \def\tR{{\tt R}}
\def\tT{{\tt T}}
\def\wt{\widetilde} \def\ov{\overline}
\def\rd{{\rm d}} \def\rS{{\rm S}} 
\def\rExt{\rm{Ext}}
\def\rInt{\rm{Int}} \def\rSupp{\rm{Supp}}

\def\onwl{\operatornamewithlimits}


\makeatletter
 \def\fps@figure{htbp}
\makeatother

\title{\bf The hard-core model on $\bbZ^3$\\ and Kepler's conjecture}

\author{\bf A. Mazel$^1$, I. Stuhl$^2$, Y. Suhov$^{2,3}$}

\date{}
\footnotetext{

\noindent
$^1$ AMC Health, New York, NY, USA;\;\;
$^2$ Math Dept, Penn State University, PA, USA;\;\;
$^3$ DPMMS, University of Cambridge and St John's College, Cambridge, UK.}

\maketitle

\begin{abstract}

We study the hard-core model of statistical mechanics on a unit cubic 
lattice $\bbZ^3$, which is intrinsically related to the sphere-packing problem  for spheres with centers in $\bbZ^3$. 
The model is defined by the sphere diameter $D>0$ which is interpreted as a Euclidean exclusion distance between point particles located at spheres centers. The second parameter of the underlying model is the particle fugacity 
$u$. For $u>1$ the ground states of the model are given by the dense-packings of the spheres. The identification of 
such dense-packings is a considerable challenge, and we solve it for $D^2=2, 3, 4, 5, 6, 8, 9, 10, 11, 12$ as well as for  
$D^2=2\ell^2$, where $\ell\in\bbN$. For the former family of values of $D^2$ our proofs are self-contained.
For $D^2=2\ell^2$ our results are based on the
proof of Kepler's conjecture. Depending on the value of $D^2$, we encounter three physically 
distinct situations: (i) finitely many periodic ground states, (ii) countably many layered periodic ground 
states and (iii) countably many not necessarily layered periodic ground states. For the first two cases we use 
the Pirogov-Sinai theory and identify the corresponding periodic Gibbs distributions for $D^2=2,3,5,8,9,10,12$ and 
$D^2=2\ell^2$, $\ell\in\bbN$, in a high-density regime $u>u_*(D^2)$, where the system is ordered and tends to 
fluctuate around some ground states. In particular, for $D^2=5$ only a finite number out of countably many layered 
periodic ground states generate pure phases. 
\end{abstract}

{\bf Significance Statement.} {Lattice hard-core models form an important class of statistical mechanical models. The complete 
understanding of these models is beyond reach, but in this work we identify the large fugacity phase diagram of periodic pure 
phases for certain exclusion diameters. In the course of the analysis we solve the sphere-packing problem on $\bbZ^3$ for the 
corresponding values of the sphere diameters. The respective solutions essentially depend on the sphere diameter, unlike the 
case of $\bbR^3$.

}

\section{Introduction. Brief description of main results.}

In this paper we consider a model of spherical particles with centers at sites of a unit cubic lattice $\bbZ^3$. A hard-core condition 
requires that the Euclidean distance $\rho(\cdot , \cdot)$ between any two particles is at least $D>0$. In other words, for this 
model a configuration $\phi=\{x_i\}$ is a collection of particles located at sites $x_i\in\bbZ^3$ such that $\rho(x_{i'}, x_{i''})
\geq D$ whenever $x_{i'} \neq  x_{i''}$. So $D$ represents the minimal allowed distance. In order to emphasize the hard-core 
requirement we frequently call such $\phi$ an admissible (or $D$-admissible) configuration. 

The energy of a finite configuration $\phi$ is given by the Hamiltonian 
$$
H(\phi):= - \ln(u) \cdot \sharp (\phi),
$$
where $\sharp (\phi)$ is the number of particles in $\phi$ and $u>0$ is the particle fugacity. The statistical weight of a configuration $\phi$ is 
$$
w(\phi) := e^{-H(\phi)}
$$
and the corresponding probability of $\phi$ is
$$\mu(\phi):=\frac{w(\phi)}{Z},$$ where 
$$
Z:= \sum_\phi w(\phi)
$$
is the associated partition function. Under the above definitions the model for a real value of $D^2$ is 
equivalent to the model with the nearest-from-above integer value $D^2=m^2+n^2+k^2$ where $m,n,k\in\bbZ$. It means that there exists a pair of points in $\bbZ^3$ at the distance $D$ from each other. Without loss of generality from now on we assume $D^2$ to be of the above form.

A central question regarding a statistical mechanics model is the complete phase diagram in the corresponding parameter space, defined by the pair $(u, D)$ for our model. The construction of the complete phase diagram is a paramount task, and in this paper we focus on the part of the diagram where $u \gg 1$. It is not hard to see from the definition of $\mu(\phi)$ that in such a regime a configuration with a larger amount of particles is more probable than a configuration with a fewer amount. In a sense, a configuration $\varphi$ containing the maximal possible amount of particles is the most probable one. Such configurations are known as ground states and are defined as $D$-admissible configurations $\varphi\in \bbZ^3$ for which one can't remove a finite number of particles and then replace them with a larger amount without breaking admissibility. 

It is clear that for $u > 1$ a ground state $\vphi$ must be a dense-packing configuration of spheres in $\bbZ^3$. A comparison of the sphere-packing problems in $\bbZ^3$ and $\bbR^3$ (cf. \cite{Ha1}, \cite{Ha2}) leads to the conclusion that the problem in $\bbZ^3$ is harder. In particular, an 
analog of  Kepler's conjecture on $\bbZ^3$ is not known. Nevertheless, it is fruitful to explore to which extent the
 existing results for $\bbR^3$ can be applied to the case of $\bbZ^3$ which is attempted in the current paper.

Since its inception in 1960's (see \cite{Bu}, \cite{GF}, \cite{Ga}), lattice H-C models attracted a considerable interest 
in the literature: physics (cf. \cite{NR1}, \cite{NR2}, \cite{VMDDR} and references therein) and mathematics 
(cf. \cite{JL}, \cite{MSS1}, \cite{MSS2}, \cite{MSS3}, \cite{MSS4}). We particularly note paper \cite{VMDDR} where 
some of our results have been predicted via numerical simulations. 

In the current paper the model is studied in the thermodynamic limit by considering limit Gibbs distributions \cite{Ge} 
in the entire lattice $\bbZ^3$. Correspondingly, pure phases are interpreted as extreme Gibbs distributions. The 
rigorous proofs of the results presented here can be found in \cite{MSS4}.

The dense-packing configurations of unit spheres in $\bbR^3$ are given by the unit FCC-lattice $\bbA_3$ and its 
layered modifications. Such modifications are obtained by shifting of some (possibly infinite) number of parallel 
$\bbA_2$-sub-lattices in $\bbA_3$ (layers). Therefore, if it is possible to inscribe/embed the scaled lattice $D\cdot\bbA_3$ 
into $\bbZ^3$ then any inscribed configuration is a dense-packing of spheres with diameter $D$ in $\bbZ^3$. The immediate 
question is for which $D$ such an inscription is possible. It is so if and only if $D^2=2\ell^2$ where $\ell\in \bbN$. Moreover, 
the associated layered dense-packings exist iff $\ell$ is divisible by $3$. 

A difference between $\bbR^3$ and $\bbZ^3$ cases is that, depending on arithmetic properties of $\ell$, there are 
multiple non-equivalent inscriptions $D\cdot\bbA_3\hookrightarrow\bbZ^3$. Here non-equivalent means that two inscriptions
cannot be taken into each other by a $\bbZ^3$-symmetry. The total number of inscriptions is finite and depends on
 certain congruence classes of the prime factors in the  prime decomposition of $\ell$. The larger the number of 
prime factors, the larger the number of non-equivalent inscriptions.

The above non-equivalence could be inessential from the sphere-packing perspective, but it is crucial from the 
statistical-mechanics point of view. In statistical mechanics, it is typical that a pure phase can be viewed
as a gas of `perturbations' around a ground state (a dense-packing in our case). Two non-equivalent 
inscriptions may yield ground states with different lists of associated perturbation structures. 
Consequently, not every inscription generates a pure phase. 

To construct a rigorous perturbation theory, one needs a local measure of deviation from
a dense-packing configuration. A natural candidate is the Euclidean volume $\ups (V(x,\phi ))$ of the Voronoi cell $V(x,\phi )\subset\bbR^3$ for a particle $x$ in a $D$-admissible configuration $\phi$. Among all 
admissible $\phi$ and all $x \in \phi$ one can always find Voronoi cells of a 
minimal volume, and there is only a finite number of possible 
shapes of such minimal cells modulo symmetries. Suppose that, for a given $D$,  
there exists an admissible configuration $\vphi$ where all cells $V(x,\vphi )$, $x\in\vphi$, have the minimal volume.
Then the volume $\ups (V(x,\phi ))$, $x\in\phi$, provides a convenient local measure of the packing density of
$\phi$ in a neighborhood of $x$. Consequently, any configuration containing only 
minimal Voronoi cells can be called a perfect configuration. A perfect configuration is always a ground state but 
not {\it vice versa}. If there exists at least one perfect configuration then any periodic ground state is a (periodic) 
perfect configuration. A non-periodic ground state may be non-perfect. 

Any particle $x\in\phi$ with a non-minimal Voronoi cell reduces the particle density. 
A connected component 
formed by non-minimal Voronoi cells constitutes a perturbation of the surrounding perfect configuration. The 
perturbations indicate parts of an admissible configuration $\phi$ where there is a deficiency of particles. Due to
discreteness, there exists a gap, $\delta(D)$, between the volumes of the minimal and the next to the minimal 
Voronoi cell. It turns out that the deficiency of particles in a perturbation can be lower bounded by 
$C D^{-6}\cdot \delta(D)\cdot{\tt v}$. Here $\tt v$ is the Euclidean volume of the perturbation and $C$ is an absolute 
constant. This estimate is known as the Peierls bound; for $u$ large enough ($u> u^{\ast}(D)$) it allows us 
to construct periodic pure phases using the machinery of the Pirogov-Sinai (P-S) theory \cite{PS}, \cite{Z}. 

It turns out that the minimal Voronoi cells tessellate $\bbZ^3$ only for sufficiently small $D$ while for larger $D$ one needs 
alternative approaches. For example, the arguments used in the proof of Kepler's conjecture can be utilized. 
More specifically, for $D^2=2\ell^2$, $\ell\in \bbN$, one can use the star-decompositions and the corresponding score functions $\sigma (x,\phi )$ (cf. \cite{Ha1}, \cite{Ha2}) in place of $\ups (V(x,\phi ))$. After such replacement,
the perturbation theory constructed in terms of $\ups (V(x,\phi ))$ can be repeated {\it verbatim}
in terms of $\sigma (x,\phi )$, $x\in\phi$. It leads to a unified approach covering both types of the local minimizer (a) the cases where $D$ is sufficiently small and admits a tessellation of $\bbZ^3$ with 
minimal Voronoi cells, and (b) the cases where $D^2=2\ell^2$, 
$\ell\in \bbN$.

If $\ell=2^n$, $n\in\bbN$, there exists a single inscription $D\cdot\bbA_3\hookrightarrow\bbZ^3$. A straightforward 
application of the P-S theory 
implies that for $u$ large enough this $D\cdot\bbA_3$-sub-lattice and its $\bbZ^3$-shifts generate distinct pure phases. 
Moreover, there are no other periodic pure phases. 

For $D^2= 2, 3, 8, 9, 10, 12$ the minimal Voronoi cell (which is unique up to a $\bbZ^3$-symmetry) does tessellate the 
entire $\bbZ^3$, and the corresponding perfect configurations are congruent to a certain sub-lattice of $\bbZ^3$. These 
sub-lattices are different from $D$-scaled FCC, in contrast to the case of $\bbR^3$ where a dense-packing lattice is always congruent to FCC. Furthermore, it appears that for each $D^2$ in the above list the corresponding dense-packing sub-lattices form a 
single $\bbZ^3$-symmetry class. Again, a straightforward application of the P-S theory implies that for $u$ large 
enough these sub-lattices and all their $\bbZ^3$-shifts generate distinct pure phases. Moreover, there are no other periodic 
pure phases. 

If $\ell\neq 2^n$ and $\ell$ is not divisible by $3$ then there exist at least two $\bbZ^3$-symmetry classes of 
$D\cdot\bbA_3$-sub-lattices of $\bbZ^3$. In the case of multiple  $\bbZ^3$-symmetry classes the P-S theory guarantees 
that for $u$ large enough for at least one $\bbZ^3$-symmetry class of $D\cdot\bbA_3$-sub-lattices the elements of this 
class and all their $\bbZ^3$-shifts generate pure phases. The identification of such a so-called dominant class requires 
a detailed analysis of the geometry and entropy of the corresponding perturbations which remains an open problem 
for a general $\ell$. We expect that the dominant class is always unique as it requires too much of a hidden symmetry 
to equalize free energies of perturbation gases given by geometrically different perturbations. On the contrary, 
$\bbZ^3$-symmetric perfect configurations have identical perturbation gases.

The case of $D^2=5$ reveals yet another phenomenon. Here the $\bbZ^3$-symmetry class of dense-packing 
sub-lattices is unique and each sub-lattice from this class is different from $\sqrt{5}\cdot\bbA_3$. Furthermore, each 
of these sub-lattices admits layered modifications (similarly to the $\bbR^3$-case) resulting in a continuum of 
perfect configurations, with countable many periodic perfect configurations among them. Such a case of infinite 
degeneracy of periodic ground states requires an advanced version of the P-S theory \cite{BS} for its analysis. In 
the course of the argument one needs to list the perturbations in the decreasing order of their statistical weights 
and analyze the corresponding free energies of the truncated (up to a certain order) perturbation gases. The perfect 
configurations with the minimal truncated free energy are the dominant ones provided there exist only finitely many 
of them. For $D^2=5$ there is a single perturbation (up to a $\bbZ^3$-symmetry) which identifies the dominant 
periodic ground states. Surprisingly, the dominant periodic ground states are analogous to HCP configurations in 
$\bbR^3$ rather than to FCC configurations in $\bbR^3$. As earlier, this advanced version of the P-S theory for $u$ large 
enough guarantees that each of the above-mentioned dominant periodic perfect configurations generates a pure 
phase. There are finitely many of them, and there are no other periodic pure phases. 

In the case of $D^2=6$ one has a similar infinite layered degeneracy of perfect configurations.  As for $D^2 = 5$, there is a single class of dense-packing sub-lattices. However, in contrast to $D^2 = 5$, each sub-lattice generates two rather than one continuum families of layered dense-packings.

For $D^2=2\ell^2$ where $\ell$ is divisible by $3$ one encounters both  phenomena. First, there exist at least
two $\bbZ^3$-symmetry classes of $D\cdot\bbA_3$-sub-lattices. Second, each $D\cdot\bbA_3$-sub-lattice admits a single continuum family of layered dense-packings (like the $\bbR^3$-case). We conjecture 
that the perturbation similar to the one in the case of $D^2=5$ resolves the infinite degeneracy of perfect 
configurations associated with a given $D\cdot\bbA_3$-sub-lattice. Consequently, the $D$-HCP configurations have 
the minimal truncated free energy within a given class of layered configurations. Here an additional perturbation 
analysis across the classes is required to isolate among them the dominant ones. We again expect that the dominant 
class is unique.

The cases of $D^2=4, 11$ demonstrate that the degeneracy of the ground states/perfect configurations can be even 
higher than for the cases $D^2=5, 6$ or $D^2=2\ell^2$ where $\ell$ is divisible by $3$. In analogy with the $\bbZ^2$ 
version of the model we refer to it as a sliding phenomenon. In the case of sliding there exists a continuum of non-layered perfect configurations. Moreover, there exist perturbations of an arbitrary large diameter with the corresponding excess in energy bounded by an absolute constant. For the case $D^2=4$ there are `too many' minimal Voronoi cells, and their combinations tessellate $\bbZ^3$ in numerous ways. The case of $D^2=11$ is considerably more complicated because it is the first value of $D$ for which the minimal Voronoi cells do not tessellate $\bbZ^3$. The investigation of this case requires an approach different from all other considered cases.

\medskip

The case of the remaining values of $D$ (different from all cases described above) is wide open. In our opinion, the 
specific questions 1) - 8) below may help to progress in the further study of the H-C model on $\bbZ^3$. In particular, 
tasks 1) and 2) may be regarded as steps towards a lattice version of Kepler's conjecture.

\begin{description}
\item{1)} Prove or disprove that for every $D^2$ there exists at least one  
periodic dense-packing  configuration. 
\item{2)} A stronger form: prove or disprove that for every $D^2$ there exists 
at least one dense-packing sub-lattice.
\item{3)} In the case of an affirmative answer to 2), develop a 
number-theoretical description of dense-packing sub-lattices.
\item{4)} Determine the list of values $D^2$ with sliding.
\item{5)} Analyze the structure of Gibbs measures for sliding values of $D^2$. 
\item{6)} Find an m-potential representation \cite{BS}, \cite{HS} of the H-C Hamiltonian for a given value of $D^2$. 
\item{7)} Prove some sort of Peierls bound for all non-sliding values of $D^2$.
\item{8)} Prove or disprove that the dominant class of periodic ground states is always unique.
\end{description}

In the subsequent sections 2 and 3 we present a more detailed account of our analysis. In
section 2 it is done for initial values of $D$ (or rather $D^2$) and in section 3 for $D=2\ell^2$ where $\ell\in\bbN$. All presented results are proven in 
\cite{MSS4}.

\section{Analysis of cases $2\leq D^2\leq 12$} 

In what follows, $\cS^{(D^2)}$ denotes the set of perfect configurations (PCs for short or $D$-PCs emphasizing to which 
value of $D$ they correspond) for a given value $D^2$. Recall, a PC is defined as an admissible configuration which at each occupied lattice site yields a maximal value of the corresponding local measure of the particle density at this site. In this section the role of such a local measure is played by the volume of the Voronoi cell, except for the case $D^2=11$. For each $D^2$ under 
consideration in this section there exists a PC that is a sub-lattice in $\bbZ^3$; such a PC is denoted by 
$\vphi^{(D^2)}$, with additional subscripts when there are several sub-lattices. FCC, HCP, BCC stand for the unit face-centered 
cubic lattice, hexagonal close-packed structure, body-centered cubic structure, respectively, while $D$-FCC, $D$-HCP, 
$D$-BCC indicate their $D$-scaled versions. Next, dFFC and dHCP indicate deformed versions of the above, specified in each concrete case. A shift of $\bbZ^3$ is referred to as a mesh; an $r^2$-mesh is the result of scaling by a factor of $r > 0$, with a distance between neighboring sites equal to $r$. A notion of an $r^2$-mesh is generalized straightforwardly to a periodic configuration in $\bbZ^3$ containing a sub-lattice. 

We focus on extreme periodic Gibbs distributions (EPGDs or $D$-EPGDs for short) describing the periodic pure 
phases of the model. All assertions about EPGDs in sections 2, 3 are made for $u$ large enough: $u>u_*(D^2)$ where $u_*(D^2)\in (0,\infty )$. 
The collection of all EPGDs for a given value $D^2$ and fugacity $u$ large enough is denoted by $\cE^{(D^2)}_{\rm{per}}$. We say that an {\rm{EPGD}}\ $\mu$ is generated by a {\rm{PC}} $\vphi$ if $\mu$ is obtained as the thermodynamic 
limit with boundary condition $\vphi$. In this case we say that $\mu$ and $\varphi$ are in a standard correspondence. It turns out that such a measure $\mu_\vphi$ is expressed in terms of a polymer 
expansion around PC $\vphi$ which gives detailed control over the decay of correlation in EPGD $\mu_\vphi$.

\bigskip
{\bf 2.1} $D^2=2$. The set $\cS^{(2)}$ has cardinality $\sharp\left(\cS^{(2)}\right)=2$ 
and consists of an FCC sub-lattice
$$\beal\vphi^{(2)}:=\big\{m\,(1,1,0) + n\,(1,0,1) + k\,(0,1,1) :\, m, n, k\in\bbZ\big\} = \bbA_3\ena$$
and its $\bbZ^3$-shift $\vphi^{(2)}+(1,0,0)$. Both PCs are periodic and layered and have the 
particle density  $1/2$. The layers are congruent triangular $2$-meshes lying in planes 
orthogonal to the main diagonals.  An example of such a layer is a two-dimensional triangular sub-lattice 
$\tau^{(2)}=\big\{m\,(1,-1,0)+n\,(1,0,-1):\;m,n\in\bbZ\}$. Cf. Fig. 1 (a,b).

Set $\cS^{(2)}$ consists of a single $\bbZ^3$-symmetry class. Consequently, the cardinality 
$\sharp\left(\cE^{(2)}_{\rm{per}}\right)=2$, and sets $\cS^{(2)}$ and $\cE^{(2)}_{\rm{per}}$
are in a standard correspondence. Cf. \cite{D}.

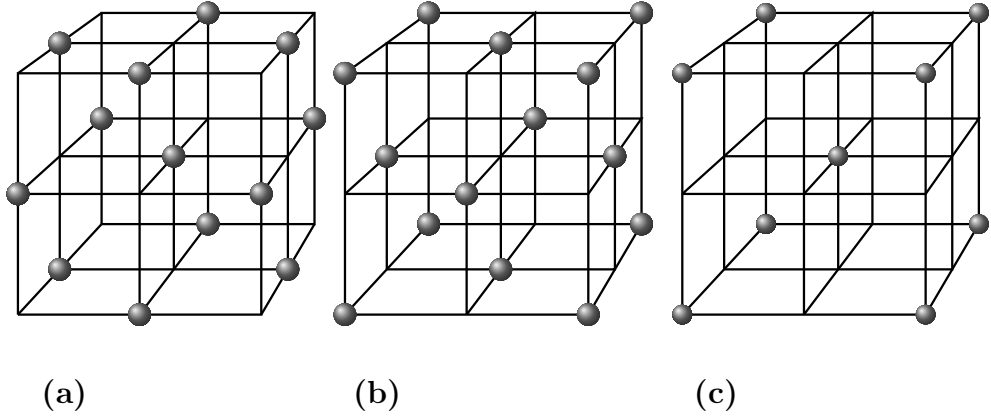
\begin{figure} \label{D-AC D^2=2,3} 
\begin{center}
\begin{tikzpicture}[scale=0.1]
\clip (-12.0, -9.0) rectangle (31.0, 35.0);

\draw [line width=0.3mm] 
(-10.0,-7.0)--(22.0,-7.0)--(22.0,25.0)--(-10.0,25.0)--(-10.0,-7.0);
\draw [line width=0.3mm] 
(1.0,5.0)--(29.0,5.0)--(29.0,33.0)--(1.0,33.0)--(1.0,5.0);

\draw [line width=0.3mm] (-10.0,-7.0)--(1.0,5.0);
\draw [line width=0.3mm] (22.0,-7.0)--(29.0,5.0);
\draw [line width=0.3mm] (22.0,25.0)--(29.0,33.0);
\draw [line width=0.3mm] (-10.0,25.0)--(1.0,33.0);

\draw [line width=0.3mm] 
(6.0,-7.0)--(6.0,25.0)--(15.0,33.0)--(15.0,5.0)--(6.0,-7.0);
\draw [line width=0.3mm] 
(-10.0,9.0)--(22.0,9.0)--(29.0,19.0)--(1.0,19.0)--(-10.0,9.0);

\draw [line width=0.3mm] (6.0,9.0)--(15.0,19.0);

\draw [line width=0.3mm] 
(-4.5,-1.0)--(25.5,-1.0)--(25.5,29.0)--(-4.5,29.0)--(-4.5,-1.0);

\draw [line width=0.3mm] (25.5,14.0)--(-4.5,14.0);
\draw [line width=0.3mm] (10.5,-1.0)--(10.5,29.0);

\foreach \pos in {(10.5,14.0),(6.0,-7.0),(6.0,25.0),(15.0,33.0),
(15.0,5.0),(6.0,-7.0),
(-4.5,-1.0),(25.5,-1.0),(25.5,29.0),(-4.5,29.0),(-4.5,-1.0),
(-10.0,9.0),(22.0,9.0),(29.0,19.0),(1.0,19.0),(-10.0,9.0)}
\shade[shading=ball, ball color=gray] \pos circle (1.5);
\end{tikzpicture}\begin{tikzpicture}[scale=0.1]
\clip (-12.0, -9.0) rectangle (31.0, 35.0);

\draw [line width=0.3mm] 
(-10.0,-7.0)--(22.0,-7.0)--(22.0,25.0)--(-10.0,25.0)--(-10.0,-7.0);
\draw [line width=0.3mm] 
(1.0,5.0)--(29.0,5.0)--(29.0,33.0)--(1.0,33.0)--(1.0,5.0);

\draw [line width=0.3mm] (-10.0,-7.0)--(1.0,5.0);
\draw [line width=0.3mm] (22.0,-7.0)--(29.0,5.0);
\draw [line width=0.3mm] (22.0,25.0)--(29.0,33.0);
\draw [line width=0.3mm] (-10.0,25.0)--(1.0,33.0);

\draw [line width=0.3mm] 
(6.0,-7.0)--(6.0,25.0)--(15.0,33.0)--(15.0,5.0)--(6.0,-7.0);
\draw [line width=0.3mm] 
(-10.0,9.0)--(22.0,9.0)--(29.0,19.0)--(1.0,19.0)--(-10.0,9.0);

\draw [line width=0.3mm] (6.0,9.0)--(15.0,19.0);

\draw [line width=0.3mm] 
(-4.5,-1.0)--(25.5,-1.0)--(25.5,29.0)--(-4.5,29.0)--(-4.5,-1.0);

\draw [line width=0.3mm] (25.5,14.0)--(-4.5,14.0);
\draw [line width=0.3mm] (10.5,-1.0)--(10.5,29.0);

\foreach \pos in {(-10.0,-7.0),(6.0,9.0),(22.0,-7.0),(22.0,25.0), (-10.0,25.0),(15.0,19.0),
(-4.5,14.0),(25.5,14.0),(10.5,-1.0),(10.5,29.0),(1.0,5.0),(29.0,5.0),(29.0,33.0),(1.0,33.0),
(1.0,5.0)}
\shade[shading=ball, ball color=gray] \pos circle (1.5);
\end{tikzpicture} \begin{tikzpicture}[scale=0.1]
\clip (-12.0, -9.0) rectangle (31.0, 35.0);

\draw [line width=0.3mm] 
(-10.0,-7.0)--(22.0,-7.0)--(22.0,25.0)--(-10.0,25.0)--(-10.0,-7.0);
\draw [line width=0.3mm] 
(1.0,5.0)--(29.0,5.0)--(29.0,33.0)--(1.0,33.0)--(1.0,5.0);

\draw [line width=0.3mm] (-10.0,-7.0)--(1.0,5.0);
\draw [line width=0.3mm] (22.0,-7.0)--(29.0,5.0);
\draw [line width=0.3mm] (22.0,25.0)--(29.0,33.0);
\draw [line width=0.3mm] (-10.0,25.0)--(1.0,33.0);

\draw [line width=0.3mm] 
(6.0,-7.0)--(6.0,25.0)--(15.0,33.0)--(15.0,5.0)--(6.0,-7.0);
\draw [line width=0.3mm] 
(-10.0,9.0)--(22.0,9.0)--(29.0,19.0)--(1.0,19.0)--(-10.0,9.0);

\draw [line width=0.3mm] (6.0,9.0)--(15.0,19.0);

\draw [line width=0.3mm] 
(-4.5,-1.0)--(25.5,-1.0)--(25.5,29.0)--(-4.5,29.0)--(-4.5,-1.0);

\draw [line width=0.3mm] (25.5,14.0)--(-4.5,14.0);
\draw [line width=0.3mm] (10.5,-1.0)--(10.5,29.0);

\foreach \pos in {(10.5,14.0),(-10.0,-7.0),(22.0,-7.0),(22.0,25.0),(-10.0,25.0),(-10.0,-7.0),
(1.0,5.0),(29.0,5.0),(29.0,33.0),(1.0,33.0),(1.0,5.0)}
\shade[shading=ball, ball color=gray] \pos circle (1.3);
\end{tikzpicture} 
\end{center}
\qquad\qquad\hskip .3cm{\bf{(a)}}\hskip 3.5cm {\bf{(b)}}\hskip 3.85cm {\bf{(c)}}

\caption{\small PCs for $D^2=2$ (the FCC sub-lattice, frame (a), and its shift, frame (b)) 
and for $D^2=3$ (a BCC sub-lattice, frame (c))}
\end{figure}

\bigskip
{\bf 2.2} $D^2=3$. The set $\cS^{(3)}$ has cardinality $\sharp\left(\cS^{(2)}\right)=4$ and 
consists of a BCC sub-lattice
$$\beal\vphi^{(3)}:=\big\{m\,(2,0,0) + n\,(0,2,0) + k\,(1,1,1) :\, m, n, k\in\bbZ\big\}\\
\ena$$
and its $\bbZ^3$-shifts $\vphi^{(3)}+(1,0,0)$, $\vphi^{(3)}+(0,1,0)$, $\vphi^{(3)}+(0,0,1)$. All PCs 
are periodic and layered and have the particle density  $1/4$.  The layers are congruent square $4$-meshes 
lying in planes orthogonal to co-ordinate axes. An example of such a layer is a square sub-lattice 
$\zeta^{(4)}:=\big\{m\,(2,0,0) + n\,(0,2,0):\;m,n\in\bbZ\big\}$.  Cf. Fig. 1 (c). 

Set $\cS^{(3)}$ consists of a single $\bbZ^3$-symmetry class. Consequently, the cardinality 
$\sharp\left(\cE^{(3)}_{\rm{per}}\right)=4$, and sets $\cS^{(3)}$ and $\cE^{(3)}_{\rm{per}}$ are in a 
standard correspondence.

\bigskip
{\bf 2.3} $D^2=4$.  The set $\cS^{(4)}$ has cardinality continuum and contains countably many 
periodic PCs. All PCs have the particle density  $1/8$ and are obtained from a lattice
$$\beal\vphi^{(4)}:=\{m\,(2,0,0) + n\,(0,2,0) + k\,(0,0,2) :\, m, n, k\in\bbZ \}=2\cdot\bbZ^3\ena$$
by shifting one- and two-dimensional meshes congruent to $2\cdot \bbZ^2$ or $2\cdot \bbZ$ in co-ordinate 
directions parallel to these meshes. A detailed formal description of set $\cS^{(4)}$ is contained in \cite{MSS4}, Theorem 3.4. This is the smallest $D^2$ exhibiting a phenomenon of sliding where 
PCs can be perturbed without a significant loss in the number of particles/gain in energy. Cf. Fig. 2. 

\begin{figure} \label{D-AC D^2=4} 
\begin{center}
\begin{tikzpicture}[scale=0.1]
\clip (-12.0, -9.0) rectangle (31.0, 35.0);

\draw [line width=0.3mm] 
(-10.0,-7.0)--(22.0,-7.0)--(22.0,25.0)--(-10.0,25.0)--(-10.0,-7.0);
\draw [line width=0.3mm] 
(1.0,5.0)--(29.0,5.0)--(29.0,33.0)--(1.0,33.0)--(1.0,5.0);

\draw [line width=0.3mm] (-10.0,-7.0)--(1.0,5.0);
\draw [line width=0.3mm] (22.0,-7.0)--(29.0,5.0);
\draw [line width=0.3mm] (22.0,25.0)--(29.0,33.0);
\draw [line width=0.3mm] (-10.0,25.0)--(1.0,33.0);

\draw [line width=0.3mm] 
(6.0,-7.0)--(6.0,25.0)--(15.0,33.0)--(15.0,5.0)--(6.0,-7.0);
\draw [line width=0.3mm] 
(-10.0,9.0)--(22.0,9.0)--(29.0,19.0)--(1.0,19.0)--(-10.0,9.0);

\draw [line width=0.3mm] (6.0,9.0)--(15.0,19.0);

\draw [line width=0.3mm] 
(-4.5,-1.0)--(25.5,-1.0)--(25.5,29.0)--(-4.5,29.0)--(-4.5,-1.0);

\draw [line width=0.3mm] (25.5,14.0)--(-4.5,14.0);
\draw [line width=0.3mm] (10.5,-1.0)--(10.5,29.0);

\foreach \pos in {(-10.0,-7.0),(22.0,-7.0),(22.0,25.0),(-10.0,25.0),
(1.0,5.0),(29.0,5.0),(29.0,33.0),(1.0,33.0),(6.0,-7.0),(6.0,25.0),
(15.0,33.0),(15.0,5.0),(-10.0,9.0),(22.0,9.0),(29.0,19.0),(1.0,19.0),
(-4.5,-1.0),(25.5,-1.0),(25.5,29.0),(-4.5,29.0),(25.5,14.0),
(-4.5,14.0),(10.5,-1.0),(10.5,29.0), (6.0,9.0),(15.0,19.0),(10.5,14.0)}
\shade[shading=ball, ball color=gray] \pos circle (1.2);
\end{tikzpicture} \begin{tikzpicture}[scale=0.1]
\clip (-12.0, -9.0) rectangle (31.0, 35.0);

\draw [line width=0.3mm] 
(-10.0,-7.0)--(22.0,-7.0)--(22.0,25.0)--(-10.0,25.0)--(-10.0,-7.0);
\draw [line width=0.3mm] 
(1.0,5.0)--(29.0,5.0)--(29.0,33.0)--(1.0,33.0)--(1.0,5.0);

\draw [line width=0.3mm] (-10.0,-7.0)--(1.0,5.0);
\draw [line width=0.3mm] (22.0,-7.0)--(29.0,5.0);
\draw [line width=0.3mm] (22.0,25.0)--(29.0,33.0);
\draw [line width=0.3mm] (-10.0,25.0)--(1.0,33.0);

\draw [line width=0.3mm] 
(6.0,-7.0)--(6.0,25.0)--(15.0,33.0)--(15.0,5.0)--(6.0,-7.0);
\draw [line width=0.3mm] 
(-10.0,9.0)--(22.0,9.0)--(29.0,19.0)--(1.0,19.0)--(-10.0,9.0);

\draw [line width=0.3mm] (6.0,9.0)--(15.0,19.0);

\draw [line width=0.3mm] 
(-4.5,-1.0)--(25.5,-1.0)--(25.5,29.0)--(-4.5,29.0)--(-4.5,-1.0);

\draw [line width=0.3mm] (25.5,14.0)--(-4.5,14.0);
\draw [line width=0.3mm] (10.5,-1.0)--(10.5,29.0);

\foreach \pos in {(22.0,-7.0),(22.0,25.0), 
(1.0,5.0),(29.0,5.0), 
(6.0,-7.0),(6.0,25.0),(15.0,5.0), 
(22.0,9.0),(29.0,19.0),(1.0,19.0), 
(-4.5,-1.0),(25.5,-1.0),(25.5,29.0),(-4.5,29.0),(-4.5,-1.0),(25.5,14.0),
(-4.5,14.0),(10.5,-1.0),(10.5,29.0), (6.0,9.0),(15.0,19.0),(10.5,14.0)}
\shade[shading=ball, ball color=gray] \pos circle (1.2);

\foreach \pos in {(-10.0,17.0),(-10.0,1.0),(8.0,33.0),(22.0,33.0)}
\shade[shading=ball, ball color=white] \pos circle (1.2);
\end{tikzpicture} \end{center}

\caption{\small PCs for $D^2=4$: a $2\bbZ^3$ sub-lattice (a), and the result of 
shifting $2$ one-dimensional meshes (b).}
\end{figure}
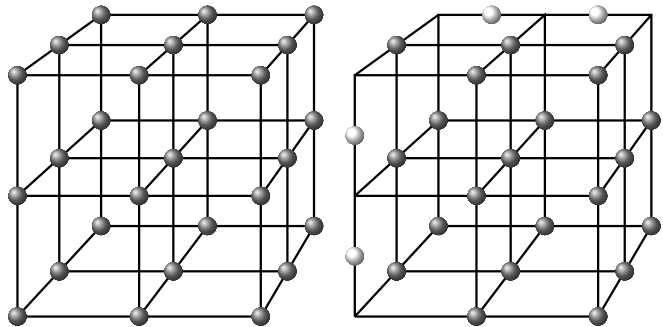

{\bf 2.4} $D^2=5$. The set $\cS^{(5)}$ has cardinality continuum and contains countably many 
periodic PCs. For instance, the AC
$$\vphi^{(5)}:=\big\{m(1,-2,1)+n(-1,-1,2)+k(1,1,1):\;m,n,k\in\bbZ\big\}$$
yields a PC that is a sub-lattice. 
 All PCs for $D^2=5$ have the particle density $1/9$ and are layered ACs where the layers are
congruent triangular $6$-meshes; an example of such a mesh is a sub-lattice $\tau^{(6)}$ in 
the plane $\rx_1+\rx_2+\rx_3=0$:
$$\tau^{(6)}=\big\{m(1,-2,1)+n(-1,-1,2):\;m,n\in\bbZ\big\}\subset\vphi^{(5)}.\eqno (1)$$ 

\begin{figure}
\begin{center}
\definecolor{blue}{gray}{0.30}
\definecolor{green}{gray}{0.55}
\definecolor{pink}{gray}{0.90}

\definecolor{white}{gray}{1.0}
\definecolor{purple}{gray}{0.4}

\begin{tikzpicture}[scale=0.7]
\path [draw=black, line width=0.4mm] (2.289, -2.077) -- (-5.688, -0.927);
\path [draw=black, line width=0.1mm] (2.857, -0.815) -- (-5.100, 0.043);
\path [draw=black, line width=0.1mm] (3.399, 0.391) -- (-4.533, 0.979);
\path [draw=black, line width=0.1mm] (3.917, 1.544) -- (-3.985, 1.883);
\path [draw=black, line width=0.1mm] (4.413, 2.648) -- (-3.455, 2.756);
\path [draw=black, line width=0.1mm] (4.889, 3.705) -- (-2.944, 3.600);
\path [draw=black, line width=0.4mm] (5.345, 4.720) -- (-2.449, 4.417);
\path [draw=black, line width=0.1mm] (2.857, -3.099) -- (-5.100, -1.751);
\path [draw=black, line width=0.1mm] (5.783, 3.684) -- (-1.970, 3.587);
\path [draw=black, line width=0.1mm] (3.399, -4.076) -- (-4.533, -2.545);
\path [draw=black, line width=0.1mm] (6.203, 2.690) -- (-1.506, 2.783);
\path [draw=black, line width=0.1mm] (3.917, -5.010) -- (-3.985, -3.312);
\path [draw=black, line width=0.1mm] (6.608, 1.733) -- (-1.057, 2.005);
\path [draw=black, line width=0.1mm] (4.413, -5.904) -- (-3.455, -4.053);
\path [draw=black, line width=0.1mm] (6.997, 0.813) -- (-0.622, 1.251);
\path [draw=black, line width=0.1mm] (4.889, -6.761) -- (-2.944, -4.769);
\path [draw=black, line width=0.1mm] (7.372, -0.073) -- (-0.200, 0.519);
\path [draw=black, line width=0.4mm] (5.345, -7.583) -- (-2.449, -5.462);
\path [draw=black, line width=0.1mm] (5.783, -6.363) -- (-1.970, -4.512);
\path [draw=black, line width=0.1mm] (6.203, -5.190) -- (-1.506, -3.593);
\path [draw=black, line width=0.1mm] (6.608, -4.063) -- (-1.057, -2.702);
\path [draw=black, line width=0.1mm] (6.997, -2.978) -- (-0.622, -1.839);
\path [draw=black, line width=0.1mm] (7.372, -1.934) -- (-0.200, -1.002);
\path [draw=black, line width=0.3mm] (7.733, -0.927) -- (0.210, -0.190);

\path [draw=black, line width=0.4mm] (2.289, -2.077) -- (5.345, -7.583);
\path [draw=black, line width=0.1mm] (0.664, -1.843) -- (3.791, -7.160);
\path [draw=black, line width=0.1mm] (-0.824, -1.628) -- (2.354, -6.769);
\path [draw=black, line width=0.1mm] (-2.189, -1.432) -- (1.022, -6.406);
\path [draw=black, line width=0.1mm] (-3.447, -1.250) -- (-0.216, -6.069);
\path [draw=black, line width=0.1mm] (-4.610, -1.083) -- (-1.370, -5.755);
\path [draw=black, line width=0.4mm] (-5.688, -0.927) -- (-2.449, -5.462);
\path [draw=black, line width=0.1mm] (2.857, -0.815) -- (5.783, -6.363);
\path [draw=black, line width=0.1mm] (-5.100, 0.043) -- (-1.970, -4.512);
\path [draw=black, line width=0.1mm] (3.399, 0.391) -- (6.203, -5.190);
\path [draw=black, line width=0.1mm] (-4.533, 0.979) -- (-1.506, -3.593);
\path [draw=black, line width=0.1mm] (3.917, 1.544) -- (6.608, -4.063);
\path [draw=black, line width=0.1mm] (-3.985, 1.883) -- (-1.057, -2.702);
\path [draw=black, line width=0.1mm] (4.413, 2.648) -- (6.997, -2.978);
\path [draw=black, line width=0.1mm] (-3.455, 2.756) -- (-0.622, -1.839);
\path [draw=black, line width=0.1mm] (4.889, 3.705) -- (7.372, -1.934);
\path [draw=black, line width=0.1mm] (-2.944, 3.600) -- (-0.200, -1.002);
\path [draw=black, line width=0.4mm] (5.345, 4.720) -- (7.733, -0.927);
\path [draw=black, line width=0.1mm] (3.791, 4.659) -- (6.259, -0.783);
\path [draw=black, line width=0.1mm] (2.354, 4.604) -- (4.884, -0.648);
\path [draw=black, line width=0.1mm] (1.022, 4.552) -- (3.601, -0.522);
\path [draw=black, line width=0.1mm] (-0.216, 4.504) -- (2.398, -0.405);
\path [draw=black, line width=0.1mm] (-1.370, 4.459) -- (1.270, -0.294);
\path [draw=black, line width=0.3mm] (-2.449, 4.417) -- (0.210, -0.190);

\path [draw=black, line width=0.4mm] (2.289, -2.077) -- (5.345, 4.720);
\path [draw=black, line width=0.1mm] (0.664, -1.843) -- (3.791, 4.659);
\path [draw=black, line width=0.1mm] (-0.824, -1.628) -- (2.354, 4.604);
\path [draw=black, line width=0.1mm] (-2.189, -1.432) -- (1.022, 4.552);
\path [draw=black, line width=0.1mm] (-3.447, -1.250) -- (-0.216, 4.504);
\path [draw=black, line width=0.1mm] (-4.610, -1.083) -- (-1.370, 4.459);
\path [draw=black, line width=0.4mm] (-5.688, -0.927) -- (-2.449, 4.417);
\path [draw=black, line width=0.1mm] (2.857, -3.099) -- (5.783, 3.684);
\path [draw=black, line width=0.1mm] (-5.100, -1.751) -- (-1.970, 3.587);
\path [draw=black, line width=0.1mm] (3.399, -4.076) -- (6.203, 2.690);
\path [draw=black, line width=0.1mm] (-4.533, -2.545) -- (-1.506, 2.783);
\path [draw=black, line width=0.1mm] (3.917, -5.010) -- (6.608, 1.733);
\path [draw=black, line width=0.1mm] (-3.985, -3.312) -- (-1.057, 2.005);
\path [draw=black, line width=0.1mm] (4.413, -5.904) -- (6.997, 0.813);
\path [draw=black, line width=0.1mm] (-3.455, -4.053) -- (-0.622, 1.251);
\path [draw=black, line width=0.1mm] (4.889, -6.761) -- (7.372, -0.073);
\path [draw=black, line width=0.1mm] (-2.944, -4.769) -- (-0.200, 0.519);
\path [draw=black, line width=0.4mm] (5.345, -7.583) -- (7.733, -0.927);
\path [draw=black, line width=0.1mm] (3.791, -7.160) -- (6.259, -0.783);
\path [draw=black, line width=0.1mm] (2.354, -6.769) -- (4.884, -0.648);
\path [draw=black, line width=0.1mm] (1.022, -6.406) -- (3.601, -0.522);
\path [draw=black, line width=0.1mm] (-0.216, -6.069) -- (2.398, -0.405);
\path [draw=black, line width=0.1mm] (-1.370, -5.755) -- (1.270, -0.294);
\path [draw=black, line width=0.3mm] (-2.449, -5.462) -- (0.210, -0.190);

\begin{scope}[fill opacity=0.6]

\filldraw[blue] (2.378, 0.151) -- (5.884, 0.043) -- (4.100, -2.545) -- (0.715, -2.351) -- (-0.893, -4.769)
 -- (-2.449, -2.169) -- (-0.893, 0.253) -- (-2.449, 2.770) -- (0.715, 2.756) -- cycle; 

\path [draw=black, line width=0.7mm] (2.378, 0.151) -- (5.884, 0.043) -- (4.100, -2.545) -- (0.715, -2.351)
 -- (-0.893, -4.769) -- (-2.449, -2.169) -- (-0.893, 0.253) -- (-2.449, 2.770) -- (0.715, 2.756) -- cycle; 

\path [draw=black, line width=0.7mm] (4.100, -2.545) -- (-2.449, -2.169) -- (0.715, 2.756) -- cycle; 

\path [draw=black, line width=0.7mm] (2.378, 0.151) -- (0.715, -2.351) -- (-0.893, 0.253) -- cycle; 

\foreach \pos in {(2.378, 0.151),(5.884, 0.043),(4.100, -2.545),(0.715, -2.351),(-0.893, -4.769),
(-2.449, -2.169),(-0.893, 0.253),(-2.449, 2.770),(0.715, 2.756)} \shade[shading=ball, ball color=blue, opacity=1] 
\pos circle (0.15); 

\filldraw[green] (-3.985, 1.883) -- (-0.706, 1.837) -- (1.022, 4.552) -- (2.815, 1.787) -- (6.608, 1.733) 
-- (4.676, -1.083) -- (6.608, -4.063) -- (2.815, -3.794) -- (1.022, -6.406) -- (-0.706, -3.544) --
 (-3.985, -3.312) -- (-2.374, -0.783) -- cycle; 

\path [draw=black, line width=0.7mm] (-3.985, 1.883) -- (-0.706, 1.837) -- (1.022, 4.552) -- (2.815, 1.787) -- (6.608, 1.733) -- (4.676, -1.083) -- (6.608, -4.063) -- (2.815, -3.794) -- (1.022, -6.406) -- (-0.706, -3.544) -- (-3.985, -3.312) -- (-2.374, -0.783) -- cycle; 

\path [draw=black, line width=0.7mm] (-0.706, 1.837) -- (2.815, 1.787) -- (4.676, -1.083) -- (2.815, -3.794) -- (-0.706, -3.544) -- (-2.374, -0.783) -- cycle; 

\path [draw=black, line width=0.7mm] (-0.706, 1.837) -- (2.815, -3.794) -- cycle; 

\path [draw=black, line width=0.7mm] (2.815, 1.787) -- (-0.706, -3.544) -- cycle; 

\path [draw=black, line width=0.7mm] (4.676, -1.083) -- (-2.374, -0.783) -- cycle; 

\path [draw=black, line width=0.7mm] (-0.491, -2.131) -- (-4.020, -1.934) -- (-2.289, 0.813) 
-- (1.378, 0.720) -- (3.322, 3.684) 
-- (5.345, 0.619) -- (3.322, -2.344) -- (5.345, -5.532) -- (1.378, -5.190) -- cycle; 

\path [draw=black, line width=0.7mm] (-2.289, 0.813) -- (5.345, 0.619) -- (1.378, -5.190) -- cycle; 

\path [draw=black, line width=0.7mm] (-0.491, -2.131) -- (1.378, 0.720) -- (3.322, -2.344) -- cycle; 

\path [draw=black, line width=1.2mm] (-3.985, 1.883)--(-2.289, 0.813);   
\path [draw=black, line width=1.2mm] (-0.706, 1.837)--(-2.289, 0.813);   
\path [draw=black, line width=1.2mm] (-2.374, -0.783)--(-2.289, 0.813);  

\path [draw=black, line width=1.2mm] (-0.706, -3.544)--(1.378, -5.190)--(2.815, -3.794)
--(3.322, -2.344)--(1.022, -0.927)--(-0.491, -2.131)--(-0.706, -3.544);  



\path [draw=black, line width=1.2mm] (-0.491, -2.131)--(-2.374, -0.783)--(-4.020, -1.934)
--(-3.985, -3.312); 

\foreach \pos in {(1.022, -0.927),(-3.985, 1.883),(-0.706, 1.837),(1.022, 4.552),(2.815, 1.787),(6.608, 1.733),
(4.676, -1.083),(6.608, -4.063),
(2.815, -3.794),(1.022, -6.406),(-0.706, -3.544),(-3.985, -3.312),(-2.374, -0.783)} \shade[shading=ball, ball 
color=green, opacity=1] \pos circle (0.15);  

\foreach \pos in {(-0.491, -2.131),(-4.020, -1.934),(-2.289, 0.813),(1.378, 0.720),(3.322, 3.684),
(5.345, 0.619),(3.322, -2.344),
(5.345, -5.532),(1.378, -5.190)} \shade[shading=ball, ball color=pink, opacity=1] \pos circle (0.15); 

\filldraw[pink] (-0.491, -2.131) -- (-4.020, -1.934) -- (-2.289, 0.813) -- (1.378, 0.720) -- (3.322, 3.684) -- (5.345, 0.619) -- 
(3.322, -2.344) -- (5.345, -5.532) -- (1.378, -5.190) -- cycle; 

\filldraw[color=white] (-2.374,-2.545) circle (.12); 
\filldraw[color=black] (-2.374,-2.545) circle (.05);

\filldraw[color=black] (1.378,-3.220) circle (.12); 
\filldraw[color=white] (1.378,-3.220) circle (.05);

\end{scope}

\path [draw=black, line width=0.4mm] (2.289, -2.077) -- (-5.688, -0.927);
\path [draw=black, line width=0.1mm] (2.857, -0.815) -- (-5.100, 0.043);
\path [draw=black, line width=0.1mm] (3.399, 0.391) -- (-4.533, 0.979);
\path [draw=black, line width=0.1mm] (3.917, 1.544) -- (-3.985, 1.883);
\path [draw=black, line width=0.1mm] (4.413, 2.648) -- (-3.455, 2.756);
\path [draw=black, line width=0.1mm] (4.889, 3.705) -- (-2.944, 3.600);
\path [draw=black, line width=0.4mm] (5.345, 4.720) -- (-2.449, 4.417);
\path [draw=black, line width=0.1mm] (2.857, -3.099) -- (-5.100, -1.751);
\path [draw=black, line width=0.1mm] (3.399, -4.076) -- (-4.533, -2.545);
\path [draw=black, line width=0.1mm] (3.917, -5.010) -- (-3.985, -3.312);
\path [draw=black, line width=0.1mm] (4.413, -5.904) -- (-3.455, -4.053);
\path [draw=black, line width=0.1mm] (4.889, -6.761) -- (-2.944, -4.769);
\path [draw=black, line width=0.4mm] (5.345, -7.583) -- (-2.449, -5.462);

\path [draw=black, line width=0.4mm] (2.289, -2.077) -- (5.345, -7.583);
\path [draw=black, line width=0.1mm] (0.664, -1.843) -- (3.791, -7.160);
\path [draw=black, line width=0.1mm] (-0.824, -1.628) -- (2.354, -6.769);
\path [draw=black, line width=0.1mm] (-2.189, -1.432) -- (1.022, -6.406);
\path [draw=black, line width=0.1mm] (-3.447, -1.250) -- (-0.216, -6.069);
\path [draw=black, line width=0.1mm] (-4.610, -1.083) -- (-1.370, -5.755);
\path [draw=black, line width=0.4mm] (-5.688, -0.927) -- (-2.449, -5.462);
\path [draw=black, line width=0.1mm] (2.857, -0.815) -- (5.783, -6.363);
\path [draw=black, line width=0.1mm] (3.399, 0.391) -- (6.203, -5.190);
\path [draw=black, line width=0.1mm] (3.917, 1.544) -- (6.608, -4.063);
\path [draw=black, line width=0.1mm] (4.413, 2.648) -- (6.997, -2.978);
\path [draw=black, line width=0.1mm] (4.889, 3.705) -- (7.372, -1.934);
\path [draw=black, line width=0.4mm] (5.345, 4.720) -- (7.733, -0.927);

\path [draw=black, line width=0.4mm] (2.289, -2.077) -- (5.345, 4.720);
\path [draw=black, line width=0.1mm] (0.664, -1.843) -- (3.791, 4.659);
\path [draw=black, line width=0.1mm] (-0.824, -1.628) -- (2.354, 4.604);
\path [draw=black, line width=0.1mm] (-2.189, -1.432) -- (1.022, 4.552);
\path [draw=black, line width=0.1mm] (-3.447, -1.250) -- (-0.216, 4.504);
\path [draw=black, line width=0.1mm] (-4.610, -1.083) -- (-1.370, 4.459);
\path [draw=black, line width=0.4mm] (-5.688, -0.927) -- (-2.449, 4.417);
\path [draw=black, line width=0.1mm] (2.857, -3.099) -- (5.783, 3.684);
\path [draw=black, line width=0.1mm] (3.399, -4.076) -- (6.203, 2.690);
\path [draw=black, line width=0.1mm] (3.917, -5.010) -- (6.608, 1.733);
\path [draw=black, line width=0.1mm] (4.413, -5.904) -- (6.997, 0.813);
\path [draw=black, line width=0.1mm] (4.889, -6.761) -- (7.372, -0.073);
\path [draw=black, line width=0.4mm] (5.345, -7.583) -- (7.733, -0.927);
\end{tikzpicture}
\end{center}
{\caption{\small A dHCP PC for $D^2=5$. Three different shades of gray represent three layers: in plane $\rx_1+\rx_2+\rx_3=-3$ (dark gray), $\rx_1+\rx_2+\rx_3=0$ (mid-gray), $\rx_1+\rx_2+\rx_3=3$ (light gray). It specifies the triangular sub-lattice $\tau^{(6)}$ and two neighboring triangular $6$-meshes. The background cubic lattice is $3\cdot\bbZ^3$. Thick black edges have squared length $5$; they connect occupied sites of sub-lattice $\tau^{(6)}$ with those of the neighboring meshes (not all such edges are shown). Furthermore, these thick black edges highlight fragments of stacks of octaherdons and stacks of tetrahedrons. The circles indicate the positions of inserted 
particles in dominant excitations of order $2$: an inserted particle is at the center of a 
common triangular face of two adjacent octahedrons (not all such positions are shown).}}
\end{figure}
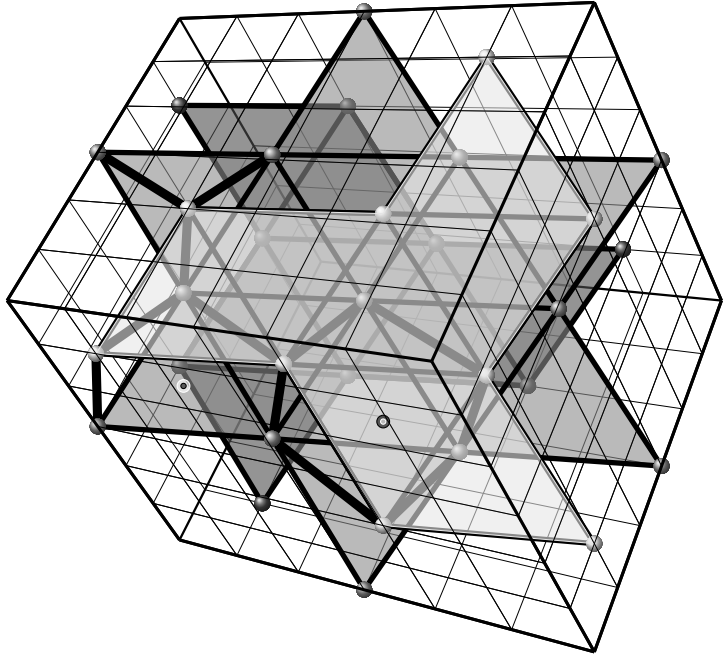

The planes containing the layers in a given PC are orthogonal to one of $4$ main diagonals 
in $\bbR^3$, and the distance between two neighboring planes equals $\sqrt 3$.
In every PC, the minimal squared distance between occupied sites equals $5$, which is achieved 
between sites in neighboring meshes. Also, the vertices in a mesh are projected to the centers 
of triangles in two neighboring meshes. The last property resembles the layered structure 
in the family of dense-packing configurations in $\bbR^3$; cf. \cite{Ha1,Ha2}. For that reason 
the corresponding PC on $\bbZ^3$ are referred to as a deformed FCC. The deformation is 
revealed by the fact that $6$ nearest neighbors of a site are at distance $\sqrt{5}$ and the 
remaining $6$ are at distance $\sqrt{6}$ from it. Likewise, continuing the $\bbR^3$-analogy, 
we define deformed HCP-configurations. In brief, we refer to dFCCs and dHCPs. For instance, 
the above sub-lattice $\vphi^{(5)}$ is a dFCC, and all other PCs that are sub-lattices are also 
dFCCs. In both dFCCs and dHCPs the triangular layers are as in $6$-FCC- and 
$6$-HCP-configurations in $\bbR^3$ but, contrary to $\bbR^3$, these layers are at a shorter 
distance from each other ($\sqrt 3$ instead of $\sqrt 4$).

The dFCCs and dHCPs, like their standard FCC and HCP counterparts, contain tetrahedrons and 
octahedrons; their edges have squared lengths $6$ (for vertices lying in the same mesh)
and $5$ (for vertices lying in neighboring meshes).

The dominant PGS-class consists of $72$ dHCPs ($18$ for each main diagonal). Consequently, 
$\sharp\left(\cE^{(5)}_{\rm{per}}\right)=72$, every dHCP generates an EPGD, 
and every EPGD\ is generated by a dHCP. 

The local excitation which distinguishes between dominant and non-dominant PGSs occurs when a particle is inserted in the center of a $6$-triangle forming a 
common face 
of two octahedrons. Such an inserted particle repels exactly three particles at the vertices of the triangle. 
See Fig. 3 where these insertions are marked by circles. Such an excitation does not occur in dFCCs, and in a dHCP it has the maximal density among all PGSs.

\bigskip
{\bf 2.5} $D^2=6$. The set $\cS^{(6)}$ has cardinality continuum and contains countably many 
periodic PCs. All PCs have the particle density $1/12$ and are layered ACs of one of two types,
(I) or (II) described below.

PCs of type (I) contain parallel layers which are congruent triangular $6$-meshes; an example of such a mesh is the triangular lattice $\tau^{(6)}$ (see (1)). Correspondingly,  the sub-lattices
$$\vphi^{(6)}_i := \big\{\tau^{(6)}+k\,\by_i:\;k\in\bbZ\big\}$$
are PCs for the following choices 
$$\by_1=(2,1,1),\;\;\by_2=(1,2,1)\;\hbox{ or }\;\by_3=(1,1,2).$$
As with $D^2=5$, the meshes in a PC of type (I) are 
lying in parallel planes orthogonal to one of the main diagonals in $\bbZ^3$;
however, the distance between neighboring planes is now $4/{\sqrt 3}$.
The neighboring meshes in a PC are arranged in such a way that an occupied site 
in a mesh one is projected to a point lying on an edge of 
the neighboring mesh and divides it at ratio $1:2$. 

The mesh $\lam_+$ neighboring $\tau^{(6)}$ and lying in the plane 
$\rx_1+\rx_2+\rx_3=4$ must be of the form $\lam_+=\lam_{+,i}=\tau^{(6)}+\by_i$, where  
the shifting vector $\by_i$ is one of $3$ vectors specified before. Furthermore,
vertex $\by_i\in\lam_{+,i}$ is projected to the point 
$\bu_i=\by_i-\left(\dfrac{4}{3},\dfrac{4}{3},\dfrac{4}{3}\right)$
which lies on an edge of $\tau^{(6)}$ incident to vertex $\bx=(0,0,0)\in\tau^{(6)}$
at distance ${\sqrt 6}/3$ from $\bx$. E.g., point $\bu_2=\left(-\dfrac{1}{3},\dfrac{2}{3},-\dfrac{1}{3}\right)$
lies on the edge $(-1,2,-1)$ emerging for $m=-1$, $n=0$ in the formula for lattice $\tau^{(6)}$.
Moreover, point $\by_i$ is the only
one in $\lam_{+,i}$ at squared distance $6$ from $\bx$. On the other hand, for the 
neighboring mesh $\lam_-=\lam_{-,i}=\tau^{(6)}-\by_i$ lying in the plane $\rx_1+\rx_2+\rx_3=-4$ 
and vertices $-\by_i$ are projected to points 
$-\by_i+\left(\dfrac{4}{3},\dfrac{4}{3},\dfrac{4}{3}\right)$ that 
lie again at distance ${\sqrt 6}/3$ from vertex $\bx=(0,0,0)$ on edges of $\tau^{(6)}$.
The above alternatives of choosing a shifting vector persist every time we move from a mesh 
to its neighbor, independently of other choices of shifting vectors. Cf. Fig. 4.

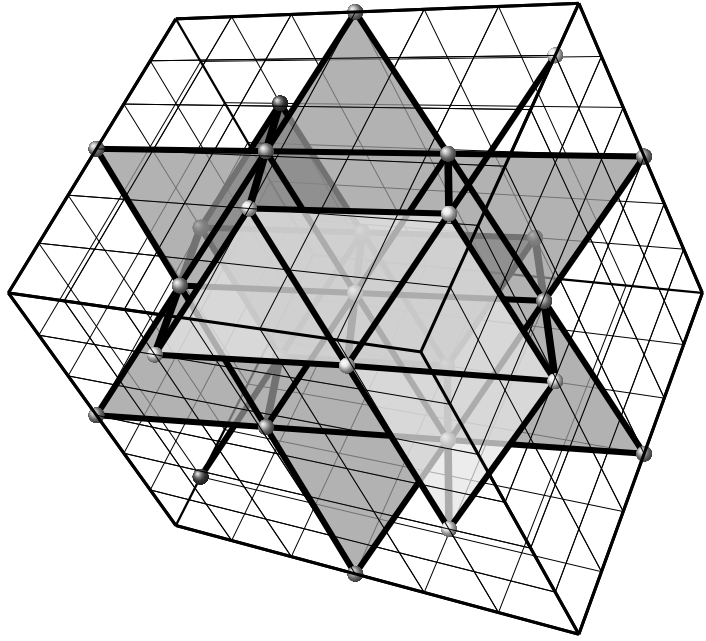
\begin{figure}
\begin{center}
\definecolor{blue}{gray}{0.30}
\definecolor{green}{gray}{0.60}
\definecolor{pink}{gray}{0.90}

\begin{tikzpicture}[scale=.68]
\path [draw=black, line width=0.4mm] (2.289, -2.077) -- (-5.688, -0.927);
\path [draw=black, line width=0.1mm] (2.857, -0.815) -- (-5.100, 0.043);
\path [draw=black, line width=0.1mm] (3.399, 0.391) -- (-4.533, 0.979);
\path [draw=black, line width=0.1mm] (3.917, 1.544) -- (-3.985, 1.883);
\path [draw=black, line width=0.1mm] (4.413, 2.648) -- (-3.455, 2.756);
\path [draw=black, line width=0.1mm] (4.889, 3.705) -- (-2.944, 3.600);
\path [draw=black, line width=0.4mm] (5.345, 4.720) -- (-2.449, 4.417);
\path [draw=black, line width=0.1mm] (2.857, -3.099) -- (-5.100, -1.751);
\path [draw=black, line width=0.1mm] (5.783, 3.684) -- (-1.970, 3.587);
\path [draw=black, line width=0.1mm] (3.399, -4.076) -- (-4.533, -2.545);
\path [draw=black, line width=0.1mm] (6.203, 2.690) -- (-1.506, 2.783);
\path [draw=black, line width=0.1mm] (3.917, -5.010) -- (-3.985, -3.312);
\path [draw=black, line width=0.1mm] (6.608, 1.733) -- (-1.057, 2.005);
\path [draw=black, line width=0.1mm] (4.413, -5.904) -- (-3.455, -4.053);
\path [draw=black, line width=0.1mm] (6.997, 0.813) -- (-0.622, 1.251);
\path [draw=black, line width=0.1mm] (4.889, -6.761) -- (-2.944, -4.769);
\path [draw=black, line width=0.1mm] (7.372, -0.073) -- (-0.200, 0.519);
\path [draw=black, line width=0.4mm] (5.345, -7.583) -- (-2.449, -5.462);
\path [draw=black, line width=0.1mm] (5.783, -6.363) -- (-1.970, -4.512);
\path [draw=black, line width=0.1mm] (6.203, -5.190) -- (-1.506, -3.593);
\path [draw=black, line width=0.1mm] (6.608, -4.063) -- (-1.057, -2.702);
\path [draw=black, line width=0.1mm] (6.997, -2.978) -- (-0.622, -1.839);
\path [draw=black, line width=0.1mm] (7.372, -1.934) -- (-0.200, -1.002);
\path [draw=black, line width=0.3mm] (7.733, -0.927) -- (0.210, -0.190);

\path [draw=black, line width=0.4mm] (2.289, -2.077) -- (5.345, -7.583);
\path [draw=black, line width=0.1mm] (0.664, -1.843) -- (3.791, -7.160);
\path [draw=black, line width=0.1mm] (-0.824, -1.628) -- (2.354, -6.769);
\path [draw=black, line width=0.1mm] (-2.189, -1.432) -- (1.022, -6.406);
\path [draw=black, line width=0.1mm] (-3.447, -1.250) -- (-0.216, -6.069);
\path [draw=black, line width=0.1mm] (-4.610, -1.083) -- (-1.370, -5.755);
\path [draw=black, line width=0.4mm] (-5.688, -0.927) -- (-2.449, -5.462);
\path [draw=black, line width=0.1mm] (2.857, -0.815) -- (5.783, -6.363);
\path [draw=black, line width=0.1mm] (-5.100, 0.043) -- (-1.970, -4.512);
\path [draw=black, line width=0.1mm] (3.399, 0.391) -- (6.203, -5.190);
\path [draw=black, line width=0.1mm] (-4.533, 0.979) -- (-1.506, -3.593);
\path [draw=black, line width=0.1mm] (3.917, 1.544) -- (6.608, -4.063);
\path [draw=black, line width=0.1mm] (-3.985, 1.883) -- (-1.057, -2.702);
\path [draw=black, line width=0.1mm] (4.413, 2.648) -- (6.997, -2.978);
\path [draw=black, line width=0.1mm] (-3.455, 2.756) -- (-0.622, -1.839);
\path [draw=black, line width=0.1mm] (4.889, 3.705) -- (7.372, -1.934);
\path [draw=black, line width=0.1mm] (-2.944, 3.600) -- (-0.200, -1.002);
\path [draw=black, line width=0.4mm] (5.345, 4.720) -- (7.733, -0.927);
\path [draw=black, line width=0.1mm] (3.791, 4.659) -- (6.259, -0.783);
\path [draw=black, line width=0.1mm] (2.354, 4.604) -- (4.884, -0.648);
\path [draw=black, line width=0.1mm] (1.022, 4.552) -- (3.601, -0.522);
\path [draw=black, line width=0.1mm] (-0.216, 4.504) -- (2.398, -0.405);
\path [draw=black, line width=0.1mm] (-1.370, 4.459) -- (1.270, -0.294);
\path [draw=black, line width=0.3mm] (-2.449, 4.417) -- (0.210, -0.190);

\path [draw=black, line width=0.4mm] (2.289, -2.077) -- (5.345, 4.720);
\path [draw=black, line width=0.1mm] (0.664, -1.843) -- (3.791, 4.659);
\path [draw=black, line width=0.1mm] (-0.824, -1.628) -- (2.354, 4.604);
\path [draw=black, line width=0.1mm] (-2.189, -1.432) -- (1.022, 4.552);
\path [draw=black, line width=0.1mm] (-3.447, -1.250) -- (-0.216, 4.504);
\path [draw=black, line width=0.1mm] (-4.610, -1.083) -- (-1.370, 4.459);
\path [draw=black, line width=0.4mm] (-5.688, -0.927) -- (-2.449, 4.417);
\path [draw=black, line width=0.1mm] (2.857, -3.099) -- (5.783, 3.684);
\path [draw=black, line width=0.1mm] (-5.100, -1.751) -- (-1.970, 3.587);
\path [draw=black, line width=0.1mm] (3.399, -4.076) -- (6.203, 2.690);
\path [draw=black, line width=0.1mm] (-4.533, -2.545) -- (-1.506, 2.783);
\path [draw=black, line width=0.1mm] (3.917, -5.010) -- (6.608, 1.733);
\path [draw=black, line width=0.1mm] (-3.985, -3.312) -- (-1.057, 2.005);
\path [draw=black, line width=0.1mm] (4.413, -5.904) -- (6.997, 0.813);
\path [draw=black, line width=0.1mm] (-3.455, -4.053) -- (-0.622, 1.251);
\path [draw=black, line width=0.1mm] (4.889, -6.761) -- (7.372, -0.073);
\path [draw=black, line width=0.1mm] (-2.944, -4.769) -- (-0.200, 0.519);
\path [draw=black, line width=0.4mm] (5.345, -7.583) -- (7.733, -0.927);
\path [draw=black, line width=0.1mm] (3.791, -7.160) -- (6.259, -0.783);
\path [draw=black, line width=0.1mm] (2.354, -6.769) -- (4.884, -0.648);
\path [draw=black, line width=0.1mm] (1.022, -6.406) -- (3.601, -0.522);
\path [draw=black, line width=0.1mm] (-0.216, -6.069) -- (2.398, -0.405);
\path [draw=black, line width=0.1mm] (-1.370, -5.755) -- (1.270, -0.294);
\path [draw=black, line width=0.3mm] (-2.449, -5.462) -- (0.210, -0.190);

\begin{scope} [fill opacity=0.75]





\filldraw[blue] (-0.432,-2.169) -- (-1.970,0.347) -- (-0.432,2.770)
-- (1.157,0.253) -- (4.499,0.151) -- (2.800,-2.351) -- (-0.432,-2.169);

\path [draw=black, line width=0.8mm] (-1.970,-4.512) -- (-0.432,-2.169) -- (-1.970,0.347) -- (-0.432,2.770)
-- (1.157,0.253) -- (4.499,0.151) -- (2.800,-2.351) -- (-0.432,-2.169);     
\path [draw=black, line width=0.8mm] (-1.970,0.347) -- (1.157,0.253);  
\path [draw=black, line width=0.8mm] (-0.432,-2.169) -- (1.157,0.253); 
\path [draw=black, line width=0.8mm] (2.800,-2.351) -- (1.157,0.253);  
 
\path [draw=black, line width=1.0mm]  (-0.706, 1.837) -- (-0.432,2.770); 
\path [draw=black, line width=1.0mm]  (-2.374, -0.783) -- (-1.970,0.347);  
\path [draw=black, line width=1.0mm]  (1.022, -0.927) -- (1.157,0.253);  
\path [draw=black, line width=1.0mm]  (4.676, -1.083) -- (4.499,0.151);  
\path [draw=black, line width=1.0mm]  (2.815, -3.794) -- (2.800,-2.351);  
\path [draw=black, line width=1.0mm] (-0.706, -3.544) -- (-0.432,-2.169);  

\foreach \pos in {(-1.970,-4.512),(-0.432,-2.169),(-0.432,2.770),(1.157,0.253),(-1.970,0.347),(2.800,-2.351),
(4.499,0.151)} \shade[shading=ball, ball color=blue, opacity=1] 
\pos circle (0.15); 

\filldraw[green] (-3.985, 1.883) -- (-0.706, 1.837) -- (1.022, 4.552) -- (2.815, 1.787) -- (6.608, 1.733) -- (4.676, -1.083) -- (6.608, -4.063) -- (2.815, -3.794) -- (1.022, -6.406) -- (-0.706, -3.544) -- (-3.985, -3.312) -- (-2.374, -0.783) -- cycle; 

\path [draw=black, line width=0.8mm] (-3.985, 1.883) -- (-0.706, 1.837) -- (1.022, 4.552) -- (2.815, 1.787) -- (6.608, 1.733) -- (4.676, -1.083) -- (6.608, -4.063) -- (2.815, -3.794) -- (1.022, -6.406) -- (-0.706, -3.544) -- (-3.985, -3.312) -- (-2.374, -0.783) -- cycle;  
\path [draw=black, line width=0.8mm] (-0.706, 1.837) -- (2.815, 1.787) -- (4.676, -1.083) -- (2.85, -3.794) -- (-0.706, -3.544) -- (-2.374, -0.783) -- cycle;   
\path [draw=black, line width=0.8mm] (-0.706, 1.837) -- (2.815, -3.794);   
\path [draw=black, line width=0.8mm] (2.815, 1.787) -- (-0.706, -3.544);   
\path [draw=black, line width=0.8mm] (4.676, -1.083) -- (-2.374, -0.783);  

\path [draw=black, line width=1.0mm]  (2.834,0.619) -- (2.815, 1.787);       
\path [draw=black, line width=1.0mm]  (4.889,-2.648) -- (4.676, -1.083);     
\path [draw=black, line width=1.0mm]  (-2.857,-2.131) -- (-2.374, -0.783);  
\path [draw=black, line width=1.0mm]  (-1.034,0.720) -- (-0.706, 1.837);     
   
\path [draw=black, line width=1.0mm] (2.834,-5.532) -- (2.815, -3.794);       
\path [draw=black, line width=1.0mm]  (0.861,-2.344) -- (1.022, -0.927);       

\foreach \pos in {(1.022, -0.927),(-3.985, 1.883),(-0.706, 1.837),(1.022, 4.552),(2.815, 1.787),(6.608, 1.733),(4.676, -1.083),(6.608, -4.063),(2.815, -3.794),(1.022, -6.406),(-0.706, -3.544),(-3.985, -3.312),(-2.374, -0.783)} \shade[shading=ball, ball color=green, opacity=1] \pos circle (0.15);  

\filldraw[pink] (2.834,0.619) -- (4.889,-2.648) -- (2.834,-5.532) -- (0.861,-2.344) -- (-2.857,-2.131) -- (-1.034,0.720) -- (2.834,0.619) ;

\path [draw=black, line width=0.8mm] (4.889,3.705) 
-- (2.834,0.619) -- (4.889,-2.648)
-- (2.834,-5.532) -- (0.861,-2.344) -- (-2.857,-2.131) -- (-1.034,0.720) -- (2.834,0.619) ;  
\path [draw=black, line width=0.8mm] (0.861,-2.344) -- (-1.034,0.720);    
\path [draw=black, line width=0.8mm] (0.861,-2.344) -- (2.834,0.619);     
\path [draw=black, line width=0.8mm] (0.861,-2.344) -- (4.889,-2.648);   

\foreach \pos in {(4.889,3.705),(0.861,-2.344),(2.834,0.619),(4.889,-2.648),(2.834,-5.532),(-2.857,-2.131),(-1.034,0.720)} 
\shade[shading=ball, ball color=pink, opacity=1] \pos circle (0.15); 


\end{scope}

\path [draw=black, line width=0.4mm] (2.289, -2.077) -- (-5.688, -0.927);
\path [draw=black, line width=0.1mm] (2.857, -0.815) -- (-5.100, 0.043);
\path [draw=black, line width=0.1mm] (3.399, 0.391) -- (-4.533, 0.979);
\path [draw=black, line width=0.1mm] (3.917, 1.544) -- (-3.985, 1.883);
\path [draw=black, line width=0.1mm] (4.413, 2.648) -- (-3.455, 2.756);
\path [draw=black, line width=0.1mm] (4.889, 3.705) -- (-2.944, 3.600);
\path [draw=black, line width=0.4mm] (5.345, 4.720) -- (-2.449, 4.417);
\path [draw=black, line width=0.1mm] (2.857, -3.099) -- (-5.100, -1.751);
\path [draw=black, line width=0.1mm] (3.399, -4.076) -- (-4.533, -2.545);
\path [draw=black, line width=0.1mm] (3.917, -5.010) -- (-3.985, -3.312);
\path [draw=black, line width=0.1mm] (4.413, -5.904) -- (-3.455, -4.053);
\path [draw=black, line width=0.1mm] (4.889, -6.761) -- (-2.944, -4.769);
\path [draw=black, line width=0.4mm] (5.345, -7.583) -- (-2.449, -5.462);

\path [draw=black, line width=0.4mm] (2.289, -2.077) -- (5.345, -7.583);
\path [draw=black, line width=0.1mm] (0.664, -1.843) -- (3.791, -7.160);
\path [draw=black, line width=0.1mm] (-0.824, -1.628) -- (2.354, -6.769);
\path [draw=black, line width=0.1mm] (-2.189, -1.432) -- (1.022, -6.406);
\path [draw=black, line width=0.1mm] (-3.447, -1.250) -- (-0.216, -6.069);
\path [draw=black, line width=0.1mm] (-4.610, -1.083) -- (-1.370, -5.755);
\path [draw=black, line width=0.4mm] (-5.688, -0.927) -- (-2.449, -5.462);
\path [draw=black, line width=0.1mm] (2.857, -0.815) -- (5.783, -6.363);
\path [draw=black, line width=0.1mm] (3.399, 0.391) -- (6.203, -5.190);
\path [draw=black, line width=0.1mm] (3.917, 1.544) -- (6.608, -4.063);
\path [draw=black, line width=0.1mm] (4.413, 2.648) -- (6.997, -2.978);
\path [draw=black, line width=0.1mm] (4.889, 3.705) -- (7.372, -1.934);
\path [draw=black, line width=0.4mm] (5.345, 4.720) -- (7.733, -0.927);

\path [draw=black, line width=0.4mm] (2.289, -2.077) -- (5.345, 4.720);
\path [draw=black, line width=0.1mm] (0.664, -1.843) -- (3.791, 4.659);
\path [draw=black, line width=0.1mm] (-0.824, -1.628) -- (2.354, 4.604);
\path [draw=black, line width=0.1mm] (-2.189, -1.432) -- (1.022, 4.552);
\path [draw=black, line width=0.1mm] (-3.447, -1.250) -- (-0.216, 4.504);
\path [draw=black, line width=0.1mm] (-4.610, -1.083) -- (-1.370, 4.459);
\path [draw=black, line width=0.4mm] (-5.688, -0.927) -- (-2.449, 4.417);
\path [draw=black, line width=0.1mm] (2.857, -3.099) -- (5.783, 3.684);
\path [draw=black, line width=0.1mm] (3.399, -4.076) -- (6.203, 2.690);
\path [draw=black, line width=0.1mm] (3.917, -5.010) -- (6.608, 1.733);
\path [draw=black, line width=0.1mm] (4.413, -5.904) -- (6.997, 0.813);
\path [draw=black, line width=0.1mm] (4.889, -6.761) -- (7.372, -0.073);
\path [draw=black, line width=0.4mm] (5.345, -7.583) -- (7.733, -0.927);
\end{tikzpicture}
\end{center}
\caption{\small PCs for $D^2=6$ of type (I). Three different shades of gray represent three layers: a triangular sub-lattice $\tau^{(6)}$ (mid-gray) in the plane $\rx_1+\rx_2+\rx_3=0$ and two neighboring $6$-meshes, in the planes $\rx_1+\rx_2+\rx_3=4$ (light gray) and $\rx_1+\rx_2+\rx_3=-4$ 
(dark gray). The background cubic lattice is $3\cdot\bbZ^3$. Thick black edges have squared length $6$ and connect occupied sites in $\tau^{(6)}$ and those in the neighboring meshes (not all such connections are shown).}
\end{figure}

\begin{figure}
\begin{center}
\definecolor{blue}{gray}{0.25}
\definecolor{green}{gray}{0.60}
\definecolor{pink}{gray}{0.90}
\begin{tikzpicture}[scale=.68]
\path [draw=black, line width=0.4mm] (2.289, -2.077) -- (-5.688, -0.927);
\path [draw=black, line width=0.1mm] (2.857, -0.815) -- (-5.100, 0.043);
\path [draw=black, line width=0.1mm] (3.399, 0.391) -- (-4.533, 0.979);
\path [draw=black, line width=0.1mm] (3.917, 1.544) -- (-3.985, 1.883);
\path [draw=black, line width=0.1mm] (4.413, 2.648) -- (-3.455, 2.756);
\path [draw=black, line width=0.1mm] (4.889, 3.705) -- (-2.944, 3.600);
\path [draw=black, line width=0.4mm] (5.345, 4.720) -- (-2.449, 4.417);
\path [draw=black, line width=0.1mm] (2.857, -3.099) -- (-5.100, -1.751);
\path [draw=black, line width=0.1mm] (5.783, 3.684) -- (-1.970, 3.587);
\path [draw=black, line width=0.1mm] (3.399, -4.076) -- (-4.533, -2.545);
\path [draw=black, line width=0.1mm] (6.203, 2.690) -- (-1.506, 2.783);
\path [draw=black, line width=0.1mm] (3.917, -5.010) -- (-3.985, -3.312);
\path [draw=black, line width=0.1mm] (6.608, 1.733) -- (-1.057, 2.005);
\path [draw=black, line width=0.1mm] (4.413, -5.904) -- (-3.455, -4.053);
\path [draw=black, line width=0.1mm] (6.997, 0.813) -- (-0.622, 1.251);
\path [draw=black, line width=0.1mm] (4.889, -6.761) -- (-2.944, -4.769);
\path [draw=black, line width=0.1mm] (7.372, -0.073) -- (-0.200, 0.519);
\path [draw=black, line width=0.4mm] (5.345, -7.583) -- (-2.449, -5.462);
\path [draw=black, line width=0.1mm] (5.783, -6.363) -- (-1.970, -4.512);
\path [draw=black, line width=0.1mm] (6.203, -5.190) -- (-1.506, -3.593);
\path [draw=black, line width=0.1mm] (6.608, -4.063) -- (-1.057, -2.702);
\path [draw=black, line width=0.1mm] (6.997, -2.978) -- (-0.622, -1.839);
\path [draw=black, line width=0.1mm] (7.372, -1.934) -- (-0.200, -1.002);
\path [draw=black, line width=0.3mm] (7.733, -0.927) -- (0.210, -0.190);

\path [draw=black, line width=0.4mm] (2.289, -2.077) -- (5.345, -7.583);
\path [draw=black, line width=0.1mm] (0.664, -1.843) -- (3.791, -7.160);
\path [draw=black, line width=0.1mm] (-0.824, -1.628) -- (2.354, -6.769);
\path [draw=black, line width=0.1mm] (-2.189, -1.432) -- (1.022, -6.406);
\path [draw=black, line width=0.1mm] (-3.447, -1.250) -- (-0.216, -6.069);
\path [draw=black, line width=0.1mm] (-4.610, -1.083) -- (-1.370, -5.755);
\path [draw=black, line width=0.4mm] (-5.688, -0.927) -- (-2.449, -5.462);
\path [draw=black, line width=0.1mm] (2.857, -0.815) -- (5.783, -6.363);
\path [draw=black, line width=0.1mm] (-5.100, 0.043) -- (-1.970, -4.512);
\path [draw=black, line width=0.1mm] (3.399, 0.391) -- (6.203, -5.190);
\path [draw=black, line width=0.1mm] (-4.533, 0.979) -- (-1.506, -3.593);
\path [draw=black, line width=0.1mm] (3.917, 1.544) -- (6.608, -4.063);
\path [draw=black, line width=0.1mm] (-3.985, 1.883) -- (-1.057, -2.702);
\path [draw=black, line width=0.1mm] (4.413, 2.648) -- (6.997, -2.978);
\path [draw=black, line width=0.1mm] (-3.455, 2.756) -- (-0.622, -1.839);
\path [draw=black, line width=0.1mm] (4.889, 3.705) -- (7.372, -1.934);
\path [draw=black, line width=0.1mm] (-2.944, 3.600) -- (-0.200, -1.002);
\path [draw=black, line width=0.4mm] (5.345, 4.720) -- (7.733, -0.927);
\path [draw=black, line width=0.1mm] (3.791, 4.659) -- (6.259, -0.783);
\path [draw=black, line width=0.1mm] (2.354, 4.604) -- (4.884, -0.648);
\path [draw=black, line width=0.1mm] (1.022, 4.552) -- (3.601, -0.522);
\path [draw=black, line width=0.1mm] (-0.216, 4.504) -- (2.398, -0.405);
\path [draw=black, line width=0.1mm] (-1.370, 4.459) -- (1.270, -0.294);
\path [draw=black, line width=0.3mm] (-2.449, 4.417) -- (0.210, -0.190);

\path [draw=black, line width=0.4mm] (2.289, -2.077) -- (5.345, 4.720);
\path [draw=black, line width=0.1mm] (0.664, -1.843) -- (3.791, 4.659);
\path [draw=black, line width=0.1mm] (-0.824, -1.628) -- (2.354, 4.604);
\path [draw=black, line width=0.1mm] (-2.189, -1.432) -- (1.022, 4.552);
\path [draw=black, line width=0.1mm] (-3.447, -1.250) -- (-0.216, 4.504);
\path [draw=black, line width=0.1mm] (-4.610, -1.083) -- (-1.370, 4.459);
\path [draw=black, line width=0.4mm] (-5.688, -0.927) -- (-2.449, 4.417);
\path [draw=black, line width=0.1mm] (2.857, -3.099) -- (5.783, 3.684);
\path [draw=black, line width=0.1mm] (-5.100, -1.751) -- (-1.970, 3.587);
\path [draw=black, line width=0.1mm] (3.399, -4.076) -- (6.203, 2.690);
\path [draw=black, line width=0.1mm] (-4.533, -2.545) -- (-1.506, 2.783);
\path [draw=black, line width=0.1mm] (3.917, -5.010) -- (6.608, 1.733);
\path [draw=black, line width=0.1mm] (-3.985, -3.312) -- (-1.057, 2.005);
\path [draw=black, line width=0.1mm] (4.413, -5.904) -- (6.997, 0.813);
\path [draw=black, line width=0.1mm] (-3.455, -4.053) -- (-0.622, 1.251);
\path [draw=black, line width=0.1mm] (4.889, -6.761) -- (7.372, -0.073);
\path [draw=black, line width=0.1mm] (-2.944, -4.769) -- (-0.200, 0.519);
\path [draw=black, line width=0.4mm] (5.345, -7.583) -- (7.733, -0.927);
\path [draw=black, line width=0.1mm] (3.791, -7.160) -- (6.259, -0.783);
\path [draw=black, line width=0.1mm] (2.354, -6.769) -- (4.884, -0.648);
\path [draw=black, line width=0.1mm] (1.022, -6.406) -- (3.601, -0.522);
\path [draw=black, line width=0.1mm] (-0.216, -6.069) -- (2.398, -0.405);
\path [draw=black, line width=0.1mm] (-1.370, -5.755) -- (1.270, -0.294);
\path [draw=black, line width=0.3mm] (-2.449, -5.462) -- (0.210, -0.190);

\begin{scope} [fill opacity=0.75]

\filldraw[blue] (-0.706, 3.630) -- (1.942, 2.741)
-- (0.446, -1.999) -- (-1.970, -1.273) -- (-0.706, 3.630);

\path [draw=black, line width=0.7mm] (-0.706, 3.630)
-- (1.942, 2.741);
\path [draw=black, line width=0.7mm] (-3.985, 1.883) -- (3.601, -0.522);  
\path [draw=black, line width=0.7mm] (-1.970, -1.273) -- (-0.706, 3.630); 
\path [draw=black, line width=0.7mm] (0.446, -1.999) -- (1.942, 2.741);   
\path [draw=black, line width=0.7mm] (0.446, -1.999) -- (-1.970, -1.273); 

\path [draw=black, line width=1.1mm] (-0.706, 3.630) -- (1.876,1.733) 
-- (1.942, 2.741);  
\path [draw=black, line width=1.1mm] (-3.985, 1.883) -- (-1.742, -0.073)
-- (-1.370, 1.054);
\path [draw=black, line width=1.1mm] (-1.370, 1.054) -- (1.022, -0.927)
-- (1.157, 0.253);  
\path [draw=black, line width=1.1mm] (-1.970, -1.273) -- (0.257, -3.312)
-- (0.446, -1.999);  
\path [draw=black, line width=1.1mm] (1.157, 0.253) -- (3.687,-1.751)
-- (3.601, -0.522);
\path [draw=black, line width=1.1mm] (-2.374,-2.545) -- (-1.970, -1.273);

\foreach \pos in {(-3.985, 1.883),(-0.706, 3.630),(-1.370, 1.054),(1.942, 2.741),
(-1.970, -1.273),(1.157, 0.253),(0.446, -1.999),(3.601, -0.522)}
\shade[shading=ball, ball color=blue, opacity=1] \pos circle (0.15);

\filldraw[green] (-2.374,-2.545) -- (-1.742, -0.073) -- (1.022, -0.927)
--(0.257, -3.312) -- (-2.374,-2.545);
\filldraw[green] (1.022, -0.927) -- (1.876,1.733) -- (4.676, 0.813)
-- (3.687,-1.751) -- (1.022, -0.927);

\path [draw=black, line width=0.7mm] (-1.742, -0.073) -- (3.687,-1.751);    
\path [draw=black, line width=0.7mm] (0.257, -3.312) -- (1.876,1.733);       
\path [draw=black, line width=0.7mm] (3.687,-1.751) -- (5.783, 3.684);       
\path [draw=black, line width=0.7mm] (-2.944, -4.769) -- (-1.742, -0.073);   
\path [draw=black, line width=0.7mm] (0.257, -3.312) -- (-5.100, -1.751);    
\path [draw=black, line width=0.7mm] (7.372, -0.073) -- (1.876,1.733);       

\path [draw=black, line width=1.1mm] (1.795,0.510) -- (1.876,1.733)
-- (4.889,-0.481);     
\path [draw=black, line width=1.1mm] (4.889,-0.481) -- (4.676, 0.813);
\path [draw=black, line width=1.1mm] (0.861, -2.344) -- (1.022, -0.927)
-- (3.791, -3.220);     
\path [draw=black, line width=1.1mm] (-2.189, -1.432) -- (-1.742, -0.073)
--(0.861, -2.344);     
\path [draw=black, line width=1.1mm] (3.791, -3.220) -- (3.687,-1.751)
-- (6.608, -4.063);     
\path [draw=black, line width=1.1mm] (0.033, -4.874) -- (0.257, -3.312) 
-- (2.815, -5.654);     

\foreach \pos in {(-5.100, -1.751),(-1.742, -0.073),(1.876,1.733),(5.783, 3.684),
(-2.374,-2.545),(1.022, -0.927),(4.676, 0.813),
(-2.944, -4.769),(0.257, -3.312),(3.687,-1.751),(7.372, -0.073)}   
 \shade[shading=ball, ball color=green, opacity=1] \pos circle (0.15);  

\filldraw[pink] (1.795,0.510) -- (4.889,-0.481) 
-- (3.791, -3.220) -- (2.815, -5.654) -- (0.033, -4.874) -- (1.795,0.510);

\path [draw=black, line width=0.7mm] (-2.189, -1.432) -- (6.608, -4.063);  
\path [draw=black, line width=0.7mm] (1.795,0.510) -- (4.889,-0.481) 
-- (3.791, -3.220) -- (2.815, -5.654) -- (0.033, -4.874) -- (1.795,0.510);      

\foreach \pos in {(-2.189, -1.432),(1.795,0.510),(0.861, -2.344),(4.889,-0.481),
(0.033, -4.874),(3.791, -3.220),(2.815, -5.654),(6.608, -4.063)} 
\shade[shading=ball, ball color=pink, opacity=1] \pos circle (0.15); 
 
\end{scope}

\path [draw=black, line width=0.4mm] (2.289, -2.077) -- (-5.688, -0.927);
\path [draw=black, line width=0.1mm] (2.857, -0.815) -- (-5.100, 0.043);
\path [draw=black, line width=0.1mm] (3.399, 0.391) -- (-4.533, 0.979);
\path [draw=black, line width=0.1mm] (3.917, 1.544) -- (-3.985, 1.883);
\path [draw=black, line width=0.1mm] (4.413, 2.648) -- (-3.455, 2.756);
\path [draw=black, line width=0.1mm] (4.889, 3.705) -- (-2.944, 3.600);
\path [draw=black, line width=0.4mm] (5.345, 4.720) -- (-2.449, 4.417);
\path [draw=black, line width=0.1mm] (2.857, -3.099) -- (-5.100, -1.751);
\path [draw=black, line width=0.1mm] (3.399, -4.076) -- (-4.533, -2.545);
\path [draw=black, line width=0.1mm] (3.917, -5.010) -- (-3.985, -3.312);
\path [draw=black, line width=0.1mm] (4.413, -5.904) -- (-3.455, -4.053);
\path [draw=black, line width=0.1mm] (4.889, -6.761) -- (-2.944, -4.769);
\path [draw=black, line width=0.4mm] (5.345, -7.583) -- (-2.449, -5.462);

\path [draw=black, line width=0.4mm] (2.289, -2.077) -- (5.345, -7.583);
\path [draw=black, line width=0.1mm] (0.664, -1.843) -- (3.791, -7.160);
\path [draw=black, line width=0.1mm] (-0.824, -1.628) -- (2.354, -6.769);
\path [draw=black, line width=0.1mm] (-2.189, -1.432) -- (1.022, -6.406);
\path [draw=black, line width=0.1mm] (-3.447, -1.250) -- (-0.216, -6.069);
\path [draw=black, line width=0.1mm] (-4.610, -1.083) -- (-1.370, -5.755);
\path [draw=black, line width=0.4mm] (-5.688, -0.927) -- (-2.449, -5.462);
\path [draw=black, line width=0.1mm] (2.857, -0.815) -- (5.783, -6.363);
\path [draw=black, line width=0.1mm] (3.399, 0.391) -- (6.203, -5.190);
\path [draw=black, line width=0.1mm] (3.917, 1.544) -- (6.608, -4.063);
\path [draw=black, line width=0.1mm] (4.413, 2.648) -- (6.997, -2.978);
\path [draw=black, line width=0.1mm] (4.889, 3.705) -- (7.372, -1.934);
\path [draw=black, line width=0.4mm] (5.345, 4.720) -- (7.733, -0.927);

\path [draw=black, line width=0.4mm] (2.289, -2.077) -- (5.345, 4.720);
\path [draw=black, line width=0.1mm] (0.664, -1.843) -- (3.791, 4.659);
\path [draw=black, line width=0.1mm] (-0.824, -1.628) -- (2.354, 4.604);
\path [draw=black, line width=0.1mm] (-2.189, -1.432) -- (1.022, 4.552);
\path [draw=black, line width=0.1mm] (-3.447, -1.250) -- (-0.216, 4.504);
\path [draw=black, line width=0.1mm] (-4.610, -1.083) -- (-1.370, 4.459);
\path [draw=black, line width=0.4mm] (-5.688, -0.927) -- (-2.449, 4.417);
\path [draw=black, line width=0.1mm] (2.857, -3.099) -- (5.783, 3.684);
\path [draw=black, line width=0.1mm] (3.399, -4.076) -- (6.203, 2.690);
\path [draw=black, line width=0.1mm] (3.917, -5.010) -- (6.608, 1.733);
\path [draw=black, line width=0.1mm] (4.413, -5.904) -- (6.997, 0.813);
\path [draw=black, line width=0.1mm] (4.889, -6.761) -- (7.372, -0.073);
\path [draw=black, line width=0.4mm] (5.345, -7.583) -- (7.733, -0.927);
\end{tikzpicture}
\end{center}
\caption{\small PCs for $D^2=6$ of type (II). Three different shades of gray represent three layers: a rhombic sub-lattice $\alpha^{(8,16)}$ 
(mid-gray) in the plane $\rx_1-\rx_2=0$ and two neighboring rhombic $(8,16)$-meshes, 
in the planes $\rx_1-\rx_2=3$ (light gray) and $\rx_1-\rx_2=-3$ (dark gray). The background cubic lattice is $3\cdot\bbZ^3$. Thick black edges have squared length $6$ and connect occupied sites in $\alpha^{(8,16)}$ and those in the neighboring meshes (not all such connections are shown).
}
\end{figure}
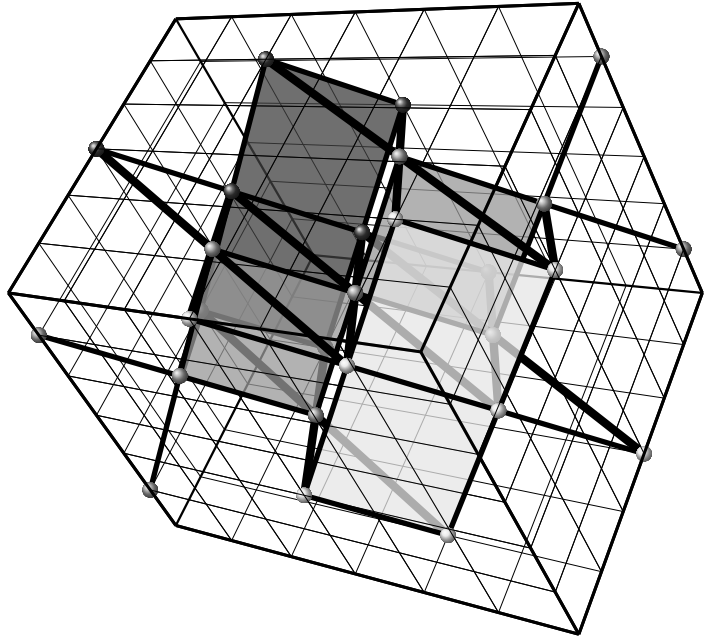

PCs of type (II) contain parallel layers which are congruent rhombic $(8,16)$-meshes;
an example of such a mesh is a rhombic lattice $\alpha^{(8,16)}$ in the plane $\rx_1-\rx_2=0$:
$$\beal\alpha^{(8,16)}=\big\{m\,(1,1,2)+n\,(1,1,-2):\;m,n\in\bbZ\big\},\ena$$
with edges of squared length $6$ and the diagonals of the rhombus of squared lengths
$8$ and $16$. Correspondingly, the sub-lattices 
$${\ov\vphi}^{(6)}_i=\big\{\alpha^{(8,16)}+k{\ov\by}_i:\;k\in\bbZ\big\}$$
yield PCs for the following equivalent choices of shifting vectors ${\ov\by}_i$:
$$\beac{\ov\by}_1=(2,-1,1)\;\;\;\hbox{ or }\;\;\;{\ov\by}_2=(1,-2,1).
\ena$$

The meshes in a PC of type (II) lie in parallel planes orthogonal to one of the non-main diagonals 
in $\bbZ^3$; the distance between neighboring planes equals $3/{\sqrt 2}$.
The neighboring meshes in such a PC are arranged so that an occupied 
site in a mesh is projected to a point lying at the middle of an edge in
a neighboring mesh.
E.g., the mesh $\nu_+$ neighboring $\alpha^{(8,16)}$ 
and lying in the plane $\rx_1-\rx_2=3$ must be of the form $\nu_+=\alpha^{(8,16)}+{\ov\by}_i$, with 
$2$ choices for the shifting vector ${\ov\by}_i$.

Similarly, the mesh $\nu_-$ neighboring $\alpha^{(8,16)}$ 
and lying in the plane $\rx_1-\rx_2=-3$ must have the same form $\alpha^{(8,16)}-{\ov\by}_i$. 

Again, these alternatives in the choice of a shifting vector persist every time we move from a mesh to its neighbor, independently of 
other choices of shifting vectors. Cf. Fig. 5.

We believe that an argument similar to the one carried out for the case $D^2=5$ works here as well, but the specific calculation isolating the dominant PGSs for $D^2=6$ has not been performed.

\bigskip
{\bf 2.6} $D^2=8$. The set $\cS^{(8)}$ has cardinality $16$ and consists of an $8$-FCC-lattice
$$\vphi^{(8)}:=\big\{m\,(2,2,0) + n\,(2,0,2) + k\,(0,2,2) :\, m, n, k\in\bbZ\big\}=2\cdot\bbA_3$$
and its $\bbZ^3$-shifts. The PCs are layered and periodic and have the particle density  $1/16$. 
The structure of layers is similar to the case $D^2=2$. Cf. Fig. 1 (a) where the 
background cubic lattice should be identified with $2\cdot\bbZ^3$. 

Set $\cS^{(8)}$ consists of a single PGS-class. Consequently, the cardinality 
$\sharp\left(\cE^{(8)}_{\rm{per}}\right)=16$, and sets $\cS^{(8)}$ and $\cE^{(8)}_{\rm{per}}$
are in a standard correspondence.

\bigskip
{\bf 2.7} $D^2=9$. The set $\cS^{(9)}$ has cardinality $120$ and consists of $6$ congruent 
sub-lattices and their $\bbZ^3$-shifts. Examples are sub-lattices $\vphi^{(9)}_{1,2}$:
$$\beal \vphi^{(9)}_1:=\Big\{m(0,3,1) + n(0,-1,3) + k(2,1,2) :\,m,n,k\in\bbZ \Big\},\\
\vphi^{(9)}_2:=\Big\{m(0,3,-1)+n(0,-1,-3)+k(2,1,-2):\;m,n,k\in\bbZ\Big\}.\ena$$
The remaining sub-lattices are obtained from $\vphi^{(9)}_i$ 
via the rotation in 
$\bbR^3$ by $\pi/2$ about the $\rx_3$- and $\rx_2$-axis, respectively. 
The PCs are layered and periodic and have the particle density $1/20$. The 
layers are congruent square $10$-meshes orthogonal to one of the co-ordinate axes.
Examples are square $10$-lattices $\zeta^{(10)}_i\subset\vphi_i^{(9)}$, $i=1,2$:
$$\beac\zeta^{(10)}_1=\big\{m\,(0,3,1) + n\,(0,-1,3):\;m,n\in\bbZ\big\},\\
\zeta^{(10)}_2=\big\{m\,(0,3,-1)+n\,(0,-1,-3):\;m,n\in\bbZ\big\}.\ena$$
Each PC is formed by parallel square $10$-meshes shifted relative to each other
by a vector of squared length $9$, where the occupied sites in a given mesh are projected 
to the centers of the squares in the neighboring meshes. For instance, in PC
$\vphi_1^{(9)}$ the neighboring meshes for $\zeta^{(10)}_1$  are $\zeta^{(10)}_1\pm (2,1,2)$, 
while in $\vphi_2^{(9)}$ the neighboring meshes for $\zeta^{(10)}_2$ are $\zeta^{(10)}_2\pm (2,1,-2)$. This structure is similar to FCC, so we 
again use the term a dFCC for configurations $\vphi\in\cS^{(9)}$. Cf. Fig. 6.

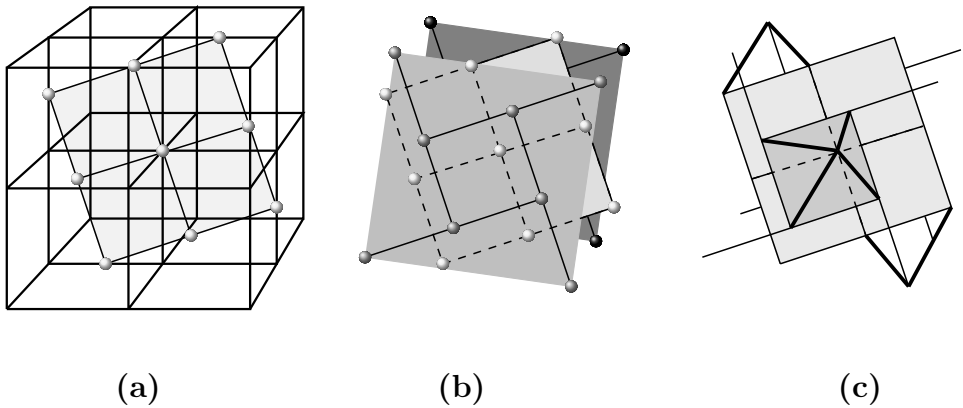
\begin{figure} \label{D-AC D^2=9} 
\begin{center}
\begin{tikzpicture}[scale=0.1] 
\clip (-12.0, -9.0) rectangle (31.0, 35.0);

\filldraw[color=gray!10] (3.0,-1.0)--(25.5,6.5)--(18.0,29.0)
--(-4.5,21.5)--(3.0,-1.0);

\draw [line width=0.3mm] 
(-10.0,-7.0)--(22.0,-7.0)--(22.0,25.0)--(-10.0,25.0)--(-10.0,-7.0);
\draw [line width=0.3mm] 
(1.0,5.0)--(29.0,5.0)--(29.0,33.0)--(1.0,33.0)--(1.0,5.0);

\draw [line width=0.3mm] (-10.0,-7.0)--(1.0,5.0);
\draw [line width=0.3mm] (22.0,-7.0)--(29.0,5.0);
\draw [line width=0.3mm] (22.0,25.0)--(29.0,33.0);
\draw [line width=0.3mm] (-10.0,25.0)--(1.0,33.0);

\draw [line width=0.3mm] 
(6.0,-7.0)--(6.0,25.0)--(15.0,33.0)--(15.0,5.0)--(6.0,-7.0);
\draw [line width=0.3mm] 
(-10.0,9.0)--(22.0,9.0)--(29.0,19.0)--(1.0,19.0)--(-10.0,9.0);

\draw [line width=0.3mm] (6.0,9.0)--(15.0,19.0);

\draw [line width=0.3mm] 
(-4.5,-1.0)--(25.5,-1.0)--(25.5,29.0)--(-4.5,29.0)--(-4.5,-1.0);

\draw [line width=0.3mm] (25.5,14.0)--(-4.5,14.0);
\draw [line width=0.3mm] (10.5,-1.0)--(10.5,29.0);

\draw [line width=0.2mm] (3.0,-1.0)--(25.5,6.5)--(18.0,29.0)
--(-4.5,21.5)--(3.0,-1.0);
\draw [line width=0.2mm] (6.75,25.25)--(14.25,2.75);
\draw [line width=0.2mm] (-0.75,10.25)--(21.75,17.25);

\foreach \pos in {(18.0,29.0),(6.75,25.25),(-4.5,21.5),
(21.75,17.25), (10.5,14.0),
(-0.75,10.25),(3.0,-1.0),(14.25,2.75),(25.5,6.5)}
\shade[shading=ball, ball color=gray!17] \pos circle (0.8);
\end{tikzpicture} \begin{tikzpicture}[scale=0.1]  
\clip (-12.0, -9.0) rectangle (31.0, 35.0);

\filldraw[color=gray] (1.365,31.0)--(26.75,27.375)
--(23.125,2.0)--(-2.25,5.625);   
  \draw [line width=0.2mm] (1.365,31.0)--(8.625,9.25);        
\draw [line width=0.2mm] (15.875,23.750)--(23.125,2.0);
\draw [line width=0.2mm] (5.0,20.125)--(26.75,27.375);
\draw [line width=0.2mm] (-2.25,5.625)--(19.50,12.875);   
\foreach \pos in {(1.365,31.0),(26.75,27.375),(23.125,2.0),(-2.25,5.625),
(5.0,20.125),(15.875,23.750),(19.50,12.875),(8.625,9.25)}
\shade[shading=ball, ball color=black] \pos circle (0.8); 

\filldraw[color=gray!25] (3.0,-1.0)--(25.5,6.5)--(18.0,29.0)
--(-4.5,21.5)--(3.0,-1.0);   
\draw [line width=0.2mm] (3.0,-1.0)--(25.5,6.5)--(18.0,29.0)
--(-4.5,21.5)--(3.0,-1.0);  

\filldraw[color=gray!50] (-3.315,27.0)--(23.75,23.125)-
-(19.875,-4.0)--(-7.25,-0.125);   

\draw [line width=0.2mm] (-3.315,27.0)--(4.375,3.75);
\draw [line width=0.2mm] (12.125,19.25)--(19.875,-4.0);
\draw [line width=0.2mm] (0.50,15.375)--(23.75,23.125);
\draw [line width=0.2mm] (-7.25,-0.125)--(16.00,7.625);

\foreach \pos in {(-3.315,27.0),(23.75,23.125),(19.875,-4.0),(-7.25,-0.125),
(12.125,19.25),(16.00,7.625),(0.50,15.375),(4.375,3.75)}
\shade[shading=ball, ball color=gray!80] \pos circle (0.8); 


\draw [line width=0.2mm, dashed] (3.0,-1.0)--(25.5,6.5)--(18.0,29.0)
--(-4.5,21.5)--(3.0,-1.0);  

\draw [line width=0.2mm, dashed] (6.75,25.25)--(14.25,2.75);     
\draw [line width=0.2mm, dashed] (-0.75,10.25)--(21.75,17.25);  

\foreach \pos in {(18.0,29.0),(6.75,25.25),(-4.5,21.5),
(21.75,17.25), (10.5,14.0),
(-0.75,10.25),(3.0,-1.0),(14.25,2.75),(25.5,6.5)}
\shade[shading=ball, ball color=gray!17] \pos circle (0.8);
\end{tikzpicture}  \begin{tikzpicture}[scale=0.1]  
\clip (-12.0, -9.0) rectangle (31.0, 35.0);


\draw [line width=0.2mm] (1.365,31.0)--(8.625,9.25);        
\draw [line width=0.2mm] (15.875,23.750)--(23.125,2.0);
\draw [line width=0.2mm] (5.0,20.125)--(26.75,27.375);
\draw [line width=0.2mm] (-2.25,5.625)--(19.50,12.875);   

\filldraw[color=gray!17] (3.0,-1.0)--(25.5,6.5)--(18.0,29.0)
--(-4.5,21.5)--(3.0,-1.0);    

\draw [line width=0.2mm] (3.0,-1.0)--(25.5,6.5)--(18.0,29.0)
--(-4.5,21.5)--(3.0,-1.0);            
\draw [line width=0.2mm] (6.75,25.25)--(14.25,2.75);     
\draw [line width=0.2mm] (-0.75,10.25)--(21.75,17.25);  

\filldraw[color=gray!40]
(12.125,19.25)--(0.50,15.375)--(4.375,3.75)--(16.00,7.625);

\draw [line width=0.2mm] (-3.315,27.0)--(4.375,3.75);      
\draw [line width=0.2mm] (12.125,19.25)--(19.875,-4.0);
\draw [line width=0.2mm] (0.50,15.375)--(23.75,23.125);
\draw [line width=0.2mm] (-7.25,-0.125)--(16.00,7.625);   

\draw [line width=0.2mm, dashed] (6.75,25.25)--(14.25,2.75);     
\draw [line width=0.2mm, dashed] (-0.75,10.25)--(21.75,17.25);  

\draw [line width=0.5mm] (19.875,-4.0)--(25.5,6.5);        
\draw [line width=0.5mm] (19.875,-4.0)--(14.25,2.75);    

\draw [line width=0.5mm] (1.365,31.0)--(-4.5,21.5);        
\draw [line width=0.5mm] (1.365,31.0)--(6.75,25.25);      

\draw [line width=0.5mm] (10.5,14.0)--(12.125,19.25);
\draw [line width=0.5mm] (10.5,14.0)--(16.00,7.625);
\draw [line width=0.5mm] (10.5,14.0)--(0.50,15.375);     
\draw [line width=0.5mm] (10.5,14.0)--(4.375,3.75);       

\end{tikzpicture} 
\end{center} 
\hskip 3cm {\bf{(a)}} \hskip 3.5cm {\bf{(b)}}  \hskip 4.5cm {\bf{(c)}} 

\caption{\small{A dFCC PC for $D^2=9$: a square sub-lattice $\zeta_1^{(10)}$ (light gray), 
and its neighboring square $10$-meshes (mid-gray, dark gray). The 
background cubic lattice in frame (a) is $4\cdot\bbZ^3$. The black edges in frame (c) have
squared length $9$; they connect
occupied sites of $\zeta_1^{(10)}$ and those in the neighboring meshes (not all such 
connections are shown).}}
\end{figure}  
 
Set $\cS^{(9)}$ consists of a single PGS-class. Consequently, the cardinality 
$\sharp\left(\cE^{(9)}_{\rm{per}}\right)=120$, and sets $\cS^{(9)}$ and $\cE^{(9)}_{\rm{per}}$
are in a standard correspondence.

\bigskip
{\bf 2.8} $D^2=10$. The set $\cS^{(10)}$ has cardinality $208$ and consists of $8$ 
congruent sub-lattices and their $\bbZ^3$-shifts. Examples are sub-lattices 
$$\beac
\vphi_1^{(10)}:=\Big\{m\,(-1,-3,4)+n\,(3,-4,1)+k\,(0 , 3, -1):\; m,n,k\in\bbZ\Big\},\\
\vphi_2^{(10)}:=\Big\{m\,(-1, 4, -3)+n\,(3, 1-4)+k\,(0, -1, 3):\; m,n,k\in\bbZ\Big\},\ena$$
obtained from each other 
via the reflection about the plane $\rx_2=\rx_3$.
The PCs are layered and periodic and have the particle density  $1/26$. The 
layers are congruent triangular $26$-meshes orthogonal to one of the main diagonals.
Examples are triangular $26$-lattices $\tau^{(26)}_i\subset\vphi_i^{(10)}$, $i=1,2$:
$$\beac\tau^{(26)}_1=\big\{m\,(-1,-3,4)+n\,(3,-4,1):\;m,n\in\bbZ\big\},\\
\tau^{(26)}_2=\big\{m\,(-1,4,-3)+n\,(3,1,-4):\;m,n\in\bbZ\big\}.\ena$$
Lattices $\tau^{(26)}_i$ lie in the plane $\rx_1+\rx_2+\rx_3=0$ and, as before, obtained 
from each other via a reflection about the plane $\rx_2=\rx_3$.
The meshes in a given PC are placed so that the minimal squared distance between 
occupied sites in neighboring meshes equals $10$, and 
the occupied sites in one mesh are projected to the centers of triangles in the
neighboring meshes. For instance, the meshes next to $\tau^{(26)}_1$ in a PC 
can be $\tau^{(26)}_1\pm (3, 1, 0)$ or $\tau^{(26)}_1\pm (1, 0, 3)$ while the meshes 
next to $\tau^{(26)}_2$ can be $\tau^{(26)}_2\pm (3,0,-1)$ or $\tau^{(26)}_2\pm (-1,3,0)$.
As before, we use the term dFCC for the PCs/PGSs constituting set $\cS^{(10)}$.
Cf. Fig. 7.

\begin{figure} \label{D-AC2 D^2=10} 
\begin{center}
\begin{tikzpicture}[scale=0.1]   
\clip (-12.0, -9.0) rectangle (31.0, 35.0);

\filldraw[gray!17] (14.5,30.5)--(0.75,30.00)--(-8.20,13.0)
--(6.6,-2.0)--(20.25,-1.50)--(27.30,15.0)--(14.5,30.5);
\draw [line width=0.2mm] (14.5,30.5)--(6.6,-2.0);
\draw [line width=0.2mm] (0.75,30.00)--(20.25,-1.50);
\draw [line width=0.2mm] (-8.20,13.0)--(27.30,15.0);
\draw [line width=0.2mm] (14.5,30.5)--(0.75,30.00)--(-8.20,13.0)
--(6.6,-2.0)--(20.25,-1.50)--(27.30,15.0)--(14.5,30.5);
\foreach \pos in {(10.5,14,0),(27.30,15.0),(0.75,30.00),(6.6,-2.0),
(14.5,30.5),(-8.20,13.0),(20.25,-1.50) }
\shade[shading=ball, ball color=gray!17] \pos circle (0.8);

\draw [line width=0.3mm] 
(-10.0,-7.0)--(22.0,-7.0)--(22.0,25.0)--(-10.0,25.0)--(-10.0,-7.0);
\draw [line width=0.3mm] 
(1.0,5.0)--(29.0,5.0)--(29.0,33.0)--(1.0,33.0)--(1.0,5.0);

\draw [line width=0.3mm] (-10.0,-7.0)--(1.0,5.0);
\draw [line width=0.3mm] (22.0,-7.0)--(29.0,5.0);
\draw [line width=0.3mm] (22.0,25.0)--(29.0,33.0);
\draw [line width=0.3mm] (-10.0,25.0)--(1.0,33.0);

\draw [line width=0.3mm] 
(6.0,-7.0)--(6.0,25.0)--(15.0,33.0)--(15.0,5.0)--(6.0,-7.0);
\draw [line width=0.3mm] 
(-10.0,9.0)--(22.0,9.0)--(29.0,19.0)--(1.0,19.0)--(-10.0,9.0);

\draw [line width=0.3mm] (6.0,9.0)--(15.0,19.0);

\draw [line width=0.3mm] 
(-4.5,-1.0)--(25.5,-1.0)--(25.5,29.0)--(-4.5,29.0)--(-4.5,-1.0);

\draw [line width=0.3mm] (25.5,14.0)--(-4.5,14.0);
\draw [line width=0.3mm] (10.5,-1.0)--(10.5,29.0);

\end{tikzpicture} \begin{tikzpicture}[scale=0.1]  
\clip (-12.0, -9.0) rectangle (31.0, 35.0);

\filldraw[gray!17] (14.5,30.5)--(0.75,30.00)--(-8.20,13.0)
--(6.6,-2.0)--(20.25,-1.50)--(27.30,15.0)--(14.5,30.5);
\draw [line width=0.2mm] (14.5,30.5)--(6.6,-2.0);
\draw [line width=0.2mm] (0.75,30.00)--(20.25,-1.50);
\draw [line width=0.2mm] (-8.20,13.0)--(27.30,15.0);
\draw [line width=0.2mm] (14.5,30.5)--(0.75,30.00)--(-8.20,13.0)
--(6.6,-2.0)--(20.25,-1.50)--(27.30,15.0)--(14.5,30.5);

\filldraw[gray!40] (12.25,26.5)--(3.0,10.5)
--(22.5,11.0)--(12.25,26.5);
\draw [line width=0.2mm] (12.25,26.5)
--(3.0,10.5)--(22.5,11.0)--(12.25,26.5);

\draw [line width=0.5mm] (10.5,14,0)--(12.25,26.5);
\draw [line width=0.5mm] (10.5,14,0)--(3.0,10.5);
\draw [line width=0.5mm] (10.5,14,0)--(22.5,11.0);

\draw [line width=0.2mm, dashed] (14.5,30.5)--(6.6,-2.0);
\draw [line width=0.2mm, dashed] (0.75,30.00)--(20.25,-1.50);
\draw [line width=0.2mm, dashed] (-8.20,13.0)--(27.30,15.0);

\foreach \pos in {(10.5,14,0),(27.30,15.0),(0.75,30.00),(6.6,-2.0),
(14.5,30.5),(-8.20,13.0),(20.25,-1.50) }
\shade[shading=ball, ball color=gray!17] \pos circle (0.8);

\foreach \pos in {(12.25,26.5),(3.0,10.5),(22.5,11.0)}
\shade[shading=ball, ball color=gray!80] \pos circle (0.8);

\end{tikzpicture} \begin{tikzpicture}[scale=0.1]. 
\clip (-12.0, -9.0) rectangle (31.0, 35.0);

\filldraw[gray!17] (14.5,30.5)--(0.75,30.00)--(-8.20,13.0)
--(6.6,-2.0)--(20.25,-1.50)--(27.30,15.0)--(14.5,30.5);

\filldraw[gray!80] (0.5,17.0)--(9.2,1)--(19.5,17.5);
\draw [line width=0.2mm, dashed] (0.5,17.0)--(9.2,1)
--(19.5,17.5)--(0.5,17.0);
\foreach \pos in {(0.5,17.0),(9.2,1),(19.5,17.5)}
\shade[shading=ball, ball color=black] \pos circle (0.8);

\draw [line width=0.2mm] (14.5,30.5)--(6.6,-2.0);
\draw [line width=0.2mm] (0.75,30.00)--(20.25,-1.50);
\draw [line width=0.2mm] (-8.20,13.0)--(27.30,15.0);
\draw [line width=0.2mm] (14.5,30.5)--(0.75,30.00)--(-8.20,13.0)
--(6.6,-2.0)--(20.25,-1.50)--(27.30,15.0)--(14.5,30.5);

\draw [line width=0.5mm] (10.5,14,0)--(0.5,17.0);
\draw [line width=0.5mm] (10.5,14,0)--(9.2,1);
\draw [line width=0.5mm] (10.5,14,0)--(19.5,17.5);

\foreach \pos in {(10.5,14,0),(27.30,15.0),(0.75,30.00),(6.6,-2.0),
(14.5,30.5),(-8.20,13.0),(20.25,-1.50) }
\shade[shading=ball, ball color=gray!17] \pos circle (0.8);

\end{tikzpicture} \end{center}
\hskip 2.75cm {\bf{(a)}} \hskip 3.75cm {\bf{(b)}}  \hskip 4cm {\bf{(c)}} 

\caption{\small A dFCC PC for $D^2=10$: a triangular sub-lattice $\tau^{(26)}_1$ (light gray)
and the neighboring triangular $26$-meshes (mid-gray, dark gray).
The background cubic lattice in frame (a) is $4\cdot\bbZ^3$. 
The thick black edges in frames (b) and (c) have squared length $10$; they connect
occupied sites in the light gray sub-lattice and those in neighboring meshes.}
\end{figure}
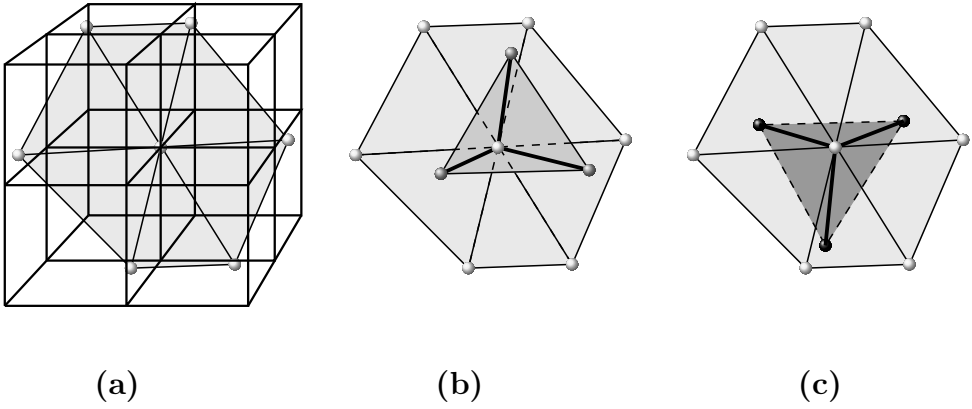  

Set $\cS^{(10)}$ consists of a single PGS-class. Consequently, the cardinality 
$\sharp\left(\cE^{(10)}_{\rm{per}}\right)=208$, and sets $\cS^{(10)}$ and $\cE^{(10)}_{\rm{per}}$
are in a standard correspondence.

\bigskip
{\bf 2.9} $D^2=11$. The set $\cS^{(11)}$ has cardinality continuum and contains countably many 
periodic PCs. All PCs have the particle density $1/32$. A continuum sub-family of $\cS^{(11)}$ can be obtained from the BCC-sub-lattice
$$\vphi^{(12)}:=\big\{m\,(4,0,0) + n\,(0,4,0) + k\,(2,2,2) :\, m, n, k\in\bbZ\big\}$$
by shifting one-dimensional meshes parallel to a given main diagonal of $\bbZ^3$. A detailed formal description of set $\cS^{(11)}$ is contained in \cite{MSS4}, Theorem 8.5. This is the second $D^2$ exhibiting a phenomenon of sliding where PCs can be perturbed without a significant loss in the number of particles/gain in energy. On the other hand this is the first case where the minimal Voronoi cell does not tessellate $\bbZ^3$ which makes it rather complicated. A corresponding investigation required a new computer assisted technique. 

\bigskip
{\bf 2.10} $D^2=12$. The set $\cS^{(12)}$ has cardinality $32$ and consists of a BCC-sub-lattice
$$\vphi^{(12)}:=\big\{m\,(4,0,0) + n\,(0,4,0) + k\,(2,2,2) :\, m, n, k\in\bbZ\big\}$$
and its $\bbZ^3$-shifts. The PCs are layered and periodic and have the particle density  $1/32$. 
Set $\cS^{(12)}$ consists of a single PGS-class. Consequently, the cardinality 
$\sharp\left(\cE^{(12)}_{\rm{per}}\right)=32$, and sets $\cS^{(12)}$ and $\cE^{(12)}_{\rm{per}}$
are in a standard correspondence. Cf. Fig. 1 (c) where the background lattice should be identified with $4\cdot\bbZ^3$.


\section{Analysis for $D^2=2\ell^2$} 

The specific examples discussed in the previous section show a broad variety of possible large-fugacity 
phase diagrams that may occur in the H-C model for different values of $D$. 
The specific methods used in those cases do not sum up to a general method applicable to all values of $D$. It may be an indication of the intrinsic difficulty which distinguishes the 
three-dimensional case from the two-dimensional one. This difficulty lies in the absence of known universal 
local optimizers identifying dense sphere packings on $\bbZ^3$ for a given value of $D$. This is in contrast 
to the case of $\,\bbZ^2$ where such a local minimizer is known (see the notion of an M-triangle in \cite{MSS1}). 
Even for the case of $\bbR^3$ (which is presumably easier than $\bbZ^3$) the search for a local optimizer took a long
 time and finally resulted in the notion of a scoring function; cf. \cite{Ha1,Ha2}, in particular, 
Definition 5.12 in \cite{Ha1}. The history of the proof of 
Kepler's conjecture illustrates the difficulty of the underlying problem. It is natural to try 
to apply the above-mentioned $\bbR^3$-results to the case of $\bbZ^3$. 

According to Kepler's conjecture,
the lattice $\bbA_3$ provides an optimal packing in $\bbR^3$. Thus, to apply this result one has to understand 
if the lattice $D\cdot \bbA_3$ can be inscribed into $\bbZ^3$ for a given $D$. An application of number-theoretical 
methods allows us to answer this question. Namely, an inscription/embedding of  $D\cdot \bbA_3$ into $\bbZ^3$ 
exists iff $D^2=2\ell^2$ where $\ell \in \bbN$. Moreover, depending on the arithmetic properties of $\ell$, there 
exists a multitude of distinct embeddings not taken into each other by $\bbZ^3$-symmetries. Nevertheless, the 
number of such embeddings is always finite. This fact solves the 
sphere-packing problem for $D^2=2\ell^2$ on $\bbZ^3$. To establish the corresponding 
structure of periodic pure phases via the PS theory one needs not only to identify dense-packings (ground states) but 
also to verify an associated Peierls bound. A certain modification \cite{MSS4} of the scoring function from 
\cite{Ha1,Ha2} can be used as the local measure of the particle density leading to the notion of a perfect configuration. 
The desired Peierls bound can be derived in the same way as in section 2. This leads to the results in 
sub-sections 3.1--3.3 below. 

These results exhaust all situations where dense-packings in $\bbZ^3$ have the same geometric structure 
as those in $\bbR^3$. Namely, they are $D\cdot\bbA_3$ meshes and (under certain additional conditions) 
their layered modifications.

\bigskip
{\bf 3.1.} Suppose that $D^2=2^{1+2n}$, corresponding to $\ell =2^n$. (It includes the values $D^2=2,8$ 
discussed before.) This case is simplified by the fact that the only 
periodic dense-packing configurations, i.e., the periodic ground states or the periodic perfect configurations
in this case are the $D$-FCC sub-lattice $D\cdot\bbA_3$ 
\noindent and its $\bbZ^3$-shifts. Thus, the only dense-packing sub-lattice forms a trivial $\bbZ^3$-symmetry 
class. Cf. Fig. 8. All other layered 
dense-packing configurations from $\bbR^3$ cannot  be inscribed in $\bbZ^3$. The analysis of PCs for 
$D^2=2^{1+2n}$ relies on \cite{Ha1,Ha2}. The definition of a contour and the 
Peierls bound are based on the notion of a scoring function from \cite{Ha1,Ha2}. 
Further details  
are discussed in \cite{MSS4}, section 9, by using a methodology developed in \cite{MSS2}. Specifically, 
cf. \cite{MSS4}, Lemmas 9.1, 9.2 and Theorems 9.1A, 9.1B.

As a result, we obtain that for $D^2=2^{1+2n}$ the entire collection of the PCs is exhausted by the set
$\sS^{(D^2)}$ consisting of $D$-FCC \ sub-lattice $2^n\cdot\bbA_3$ and its $\bbZ^3$-shifts.
The cardinality of \ $\sS^{(D^2)}$ is $2^{1+3n}$. The particle density of any PC \ 
equals $1/2^{1+3n}$. All PCs are periodic, and they form a single $\bbZ^3$-symmetry class. Consequently, the cardinality $\sharp (\sE^{(D^2)}_{\rm{per}})=2^{1+3n}$, and sets $\sS^{(D^2)}$ and $\sE^{(D^2)}_{\rm{per}}$ are in a standard correspondence. Cf. \cite{MSS4}, Theorems 9.1A, 9.1B.

\vskip -40pt
\begin{figure}[t] \label{D-AC3 D^2=5} 
\begin{center}\begin{tikzpicture}[scale=0.1]
\clip (-12.0, -9.0) rectangle (31.0, 35.0);

\draw [line width=0.3mm] 
(-10.0,-7.0)--(22.0,-7.0)--(22.0,25.0)--(-10.0,25.0)--(-10.0,-7.0);
\draw [line width=0.3mm] 
(1.0,5.0)--(29.0,5.0)--(29.0,33.0)--(1.0,33.0)--(1.0,5.0);

\draw [line width=0.3mm] (-10.0,-7.0)--(1.0,5.0);
\draw [line width=0.3mm] (22.0,-7.0)--(29.0,5.0);
\draw [line width=0.3mm] (22.0,25.0)--(29.0,33.0);
\draw [line width=0.3mm] (-10.0,25.0)--(1.0,33.0);

\draw [line width=0.3mm] 
(6.0,-7.0)--(6.0,25.0)--(15.0,33.0)--(15.0,5.0)--(6.0,-7.0);
\draw [line width=0.3mm] 
(-10.0,9.0)--(22.0,9.0)--(29.0,19.0)--(1.0,19.0)--(-10.0,9.0);

\draw [line width=0.3mm] (6.0,9.0)--(15.0,19.0);

\draw [line width=0.3mm] 
(-4.5,-1.0)--(25.5,-1.0)--(25.5,29.0)--(-4.5,29.0)--(-4.5,-1.0);

\draw [line width=0.3mm] (25.5,14.0)--(-4.5,14.0);
\draw [line width=0.3mm] (10.5,-1.0)--(10.5,29.0);

\foreach \pos in {(10.5,14.0),(6.0,25.0),(15.0,33.0),
(15.0,5.0),(6.0,-7.0),
(-4.5,-1.0),(25.5,-1.0),(25.5,29.0),(-4.5,29.0),
(-10.0,9.0),(22.0,9.0),(29.0,19.0),(1.0,19.0),(-10.0,9.0)}
\shade[shading=ball, ball color=gray] \pos circle (1.0);

\end{tikzpicture} \begin{tikzpicture}[scale=0.1] 
\clip (-12.0, -9.0) rectangle (31.0, 35.0);

\filldraw[gray] (1.0,19.0)--(-4.5,-1.0)--(15.0,5.0);   
\draw [line width=0.2mm] (1.0,19.0)--(-4.5,-1.0)
--(15.0,5.0)--(1.0,19.0);   

\filldraw[gray!17] (15.0,33.0)--(-4.5,29.0)
--(-10.0,9.0)--(6.0,-7.0)--(25.5,-1.0)--(29.0,19.0)--(15.0,33.0); 

\draw [line width=0.2mm] (15.0,33.0)--(-4.5,29.0)   
--(-10.0,9.0)--(6.0,-7.0)--(25.5,-1.0)--(29.0,19.0)--(15.0,33.0);  
\draw [line width=0.2mm] (15.0,33.0)--(6.0,-7.0);   
\draw [line width=0.2mm] (-4.5,29.0)--(25.5,-1.0);  
\draw [line width=0.2mm] (-10.0,9.0)--(29.0,19.0);  

\filldraw[gray!40] (25.5,29.0)--(22.0,9.0)--(6.0,25.0)--(25.5,29.0);   
\draw [line width=0.2mm] (25.5,29.0)--(6.0,25.0)--(22.0,9.0)--(25.5,29.0);   

\draw [line width=0.2mm, dashed] (1.0,19.0)--(-4.5,-1.0)
--(15.0,5.0)--(1.0,19.0);   

\foreach \pos in {(10.5,14.0),(15.0,33.0),(-4.5,29.0),
(-10.0,9.0),(6.0,-7.0),(25.5,-1.0),(29.0,19.0)}
\shade[shading=ball, ball color=gray!17] \pos circle (1.0);  

\foreach \pos in {(1.0,19.0),(-4.5,-1.0),(15.0,5.0)}
\shade[shading=ball, ball color=black] \pos circle (1.0); 

\foreach \pos in {(25.5,29.0),(22.0,9.0),(6.0,25.0)}
\shade[shading=ball, ball color=gray!80] \pos circle (1.0);  

\end{tikzpicture} \vskip .1cm
{\bf{(a)}}\hskip 4cm {\bf{(b)}} 
\end{center}
\begin{center}\begin{tikzpicture}[scale=0.1] 
\clip (-12.0, -9.0) rectangle (31.0, 35.0);

\draw [line width=0.3mm] 
(-10.0,-7.0)--(22.0,-7.0)--(22.0,25.0)--(-10.0,25.0)--(-10.0,-7.0);
\draw [line width=0.3mm] 
(1.0,5.0)--(29.0,5.0)--(29.0,33.0)--(1.0,33.0)--(1.0,5.0);

\draw [line width=0.3mm] (-10.0,-7.0)--(1.0,5.0);
\draw [line width=0.3mm] (22.0,-7.0)--(29.0,5.0);
\draw [line width=0.3mm] (22.0,25.0)--(29.0,33.0);
\draw [line width=0.3mm] (-10.0,25.0)--(1.0,33.0);

\draw [line width=0.3mm] 
(6.0,-7.0)--(6.0,25.0)--(15.0,33.0)--(15.0,5.0)--(6.0,-7.0);
\draw [line width=0.3mm] 
(-10.0,9.0)--(22.0,9.0)--(29.0,19.0)--(1.0,19.0)--(-10.0,9.0);

\draw [line width=0.3mm] (6.0,9.0)--(15.0,19.0);

\draw [line width=0.3mm] 
(-4.5,-1.0)--(25.5,-1.0)--(25.5,29.0)--(-4.5,29.0)--(-4.5,-1.0);

\draw [line width=0.3mm] (25.5,14.0)--(-4.5,14.0);
\draw [line width=0.3mm] (10.5,-1.0)--(10.5,29.0);

\foreach \pos in {(-10.0,-7.0),(6.0,9.0),(22.0,-7.0),(22.0,25.0), (-10.0,25.0),(15.0,19.0),
(-4.5,14.0),(25.5,14.0),(10.5,-1.0),(10.5,29.0),(1.0,5.0),(29.0,5.0),(29.0,33.0),(1.0,33.0),
(1.0,5.0)}
\shade[shading=ball, ball color=gray] \pos circle (1.0);
\end{tikzpicture} \begin{tikzpicture}[scale=0.1]  
\clip (-12.0, -9.0) rectangle (31.0, 35.0);

\filldraw[gray!17] (1.0,33.0)
--(-10.0,-7.0)--(29.0,5.0)--(1.0,33.0);    
\draw [line width=0.2mm] (1.0,33.0)
--(-10.0,-7.0)--(29.0,5.0)--(1.0,33.0);    
\draw [line width=0.2mm] (15.0,19.0)
--(-4.5,14.0)--(10.5,-1.0)-- (15.0,19.0);  

\filldraw[gray!40] (-10.0,25.0)--(22.0,-7.0)--(29.0,33.0)--(-10.0,25.0);      
\draw [line width=0.2mm] (-10.0,25.0)--(22.0,-7.0)
--(29.0,33.0)--(-10.0,25.0);   
\draw [line width=0.2mm] (6.0,9.0)--(10.5,29.0)--(25.5,14.0)--(6.0,9.0);  

\draw [line width=0.2mm, dashed] (1.0,33.0)
--(-10.0,-7.0)--(29.0,5.0)--(1.0,33.0);    
\draw [line width=0.2mm, dashed] (15.0,19.0)
--(-4.5,14.0)--(10.5,-1.0)-- (15.0,19.0);  

\foreach \pos in {(22.0,25.0),(1.0,5.0)}
\shade[shading=ball, ball color=black] \pos circle (1.0);  

\foreach \pos in {(-10.0,-7.0),(15.0,19.0),(-4.5,14.0),(10.5,-1.0),(29.0,5.0),(1.0,33.0)}
\shade[shading=ball, ball color=gray!17] \pos circle (1.0); 

\foreach \pos in {(-10.0,25.0),(22.0,-7.0),(29.0,33.0),(10.5,29.0),(25.5,14.0),(6.0,9.0)}
\shade[shading=ball, ball color=gray!80] \pos circle (1.0);  

\end{tikzpicture} 
\vskip .1cm
{\bf{(c)}}\hskip 4cm {\bf{(d)}}
\end{center}

\caption{\small An FCC-PC for $D^2=2^{1+2n}$ (frames (a,b)) and its $\bbZ^3$-shift by 
the amount $D/{\sqrt 2}$
(frames (c,d)). There are middle triangular meshes (light gray) and their neighboring meshes 
(mid-gray, dark gray), all of squared size $D^2$. The background cubic lattice in frames (a,c) is 
$(D/{\sqrt 2})\cdot\bbZ^3$.}
\end{figure}
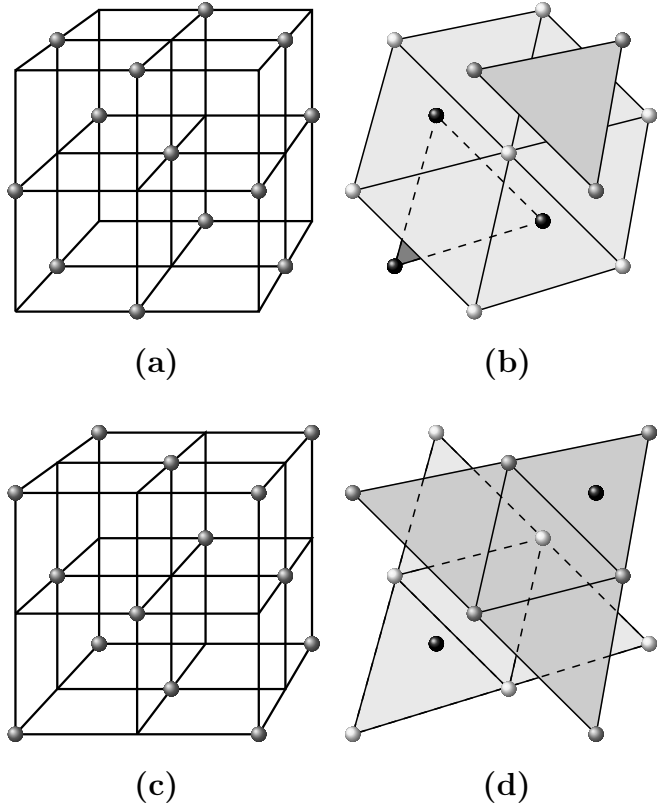  


\medskip
{\bf 3.2.} In the general case $D^2=2\ell^2$ where $\ell\in\bbN$, there still exist PCs that are
$D$-FCC sub-lattices and their $\bbZ^3$-shifts. For $\ell\neq 2^n$, such PCs 
form more than one $\bbZ^3$-symmetry class; the precise number of PC-equivalence classes and the 
symmetries within a given class depend on arithmetic properties of $\ell$. For example, 
the number of $D$-FCC sub-lattices in a class can be $4$, $6$, $8$, $12$ or $24$. A detailed analysis is given in \cite{MSS4}, section 10 (Appendix A).

Suppose that $\ell\neq 0\mod 3$ and $\ell\neq 2^n$  (the first value of $D^2$ with this property is $D^2=50$,
for $\ell =5$.)  Then the corresponding set \ $\sS^{(D^2)}$ of all PCs contains only $D$-FCC sub-lattices and their $\bbZ^3$-shifts. That is, there are no non-sub-lattice layered PCs (unlike in case $\bbR^3$). 
Set \ $\sS^{(D^2)}$ is finite and all PCs are periodic. The particle density in any \ PC \ equals $1/2\ell^3$. 
The PCs form more than one $\bbZ^3$-symmetry class.  As was said, the number of equivalence classes and 
their cardinalities depend on the rational prime decomposition of $\ell$. 
 
Furthermore, there exists $u_\star (D^2)\in (0,\infty )$ such that for $u\geq u_\star (D^2)$ there  
exists at least one dominant PGS-equivalence class. Each \ PGS \ $\vphi$ from a dominant class generates an EPGM\ $\mu_\vphi$, and every \ EPGM \ $\mu$  is generated by a \ PGS \ from 
some dominant class. Cf. \cite{MSS4}, Theorems 9.2A, 9.2B.

Specification of dominant classes for a general  $\ell\neq 0\mod 3$ and $\ell\neq 2^n$ could be done by 
studying  a `truncated free 
energy' for the ensemble of small contours (cf. \cite{MSS2,MSS3} where the H-C model was discussed on planar 
lattices). The precise answer requires a careful analysis of local `excitations' of a given PGS (where some particles are 
removed and others added, without breaking admissibility); in \cite{MSS2} it
involved a computer-assisted argument. 
We conjecture that a dominant class for $\ell\neq 0\mod 3$ is always unique.

\bigskip
{\bf 3.3.} Suppose that $D^2=2\ell^2$ where $\ell\in\bbN$ and $\ell = 0\mod 3$. (The first such value is $D^2=18$, 
for $\ell =3$.) In this case we 
encounter two phenomena. First, there exist at least two $\bbZ^3$-symmetry classes of $D\cdot\bbA_3$-sub-lattices. 
Second, each $D\cdot\bbA_3$-sub-lattice admits a single continuum family of layered dense-packings (like the 
$\bbR^3$-case). It results in a family $\sS^{(D^2)}$ of $\bbZ^3$-inscribed layered dense-packing configurations that are inherited from 
$\bbR^3$. Set $\sS^{(D^2)}$ exhausts all PCs and has cardinality continuum. The particle density in a \ PC \ equals \ 
$1/2\ell^3$. The subset $\sS^{(D^2)}_{\rm{per}}\subset\sS^{(D^2)}$ consisting of periodic layered \ PCs \ 
from \ $\sS^{(D^2)}$ is countable and exhausts all periodic \ PCs. 
Depending on the rational prime decomposition of $\ell$, the sub-lattices in $\sS^{(D^2)}_{\rm{per}}$ are 
partitioned into more than one but finitely many $\bbZ^3$-symmetry classes. Cf. \cite{MSS4}, Theorem 9.3A.

We suggest that in a sense this case is similar to the case of $D^2=5$ considered in section 2.4. Namely, there exists a local excitation of order  $2$ 
(one particle added, three removed) which has the maximal frequency of occurrence in 
HCP-type configurations. (The $D$-FCC PGSs do not have such excitations at all.) We expect that as in the case $D^2=5$ there is no other excitation of this order. If this expectation is true then in a given $\bbZ^3$-symmetry class the corresponding $D$-HCP-type configurations dominate all the remaining continuum of PCs. Nevertheless, $D$-HCP-type configurations originating from different $\bbZ^3$-symmetry classes may differ in higher order local exitations. We conjecture that this is the case and only a single $\bbZ^3$-symmetry class produces $D$-HCP-type configurations that are overall dominant.

 \vskip .5cm


{\bf Acknowledgement.} {IS and YS thank the Math Department, Penn State 
University, for support. A part of the work was carried out when IS was a 
Beaufort Visiting Fellow at St John's College, Cambridge. IS thanks St John's College, Cambridge and DPMMS, University of Cambridge, for the warm hospitality and support. YS thanks St John's College, Cambridge, for support.}

 \end{document}